\documentclass[useAMS,usenatbib]{mnras}

\usepackage{amsmath}
\usepackage{amssymb}

\usepackage{natbib}
\usepackage{graphicx}
\usepackage[T1]{fontenc}

\usepackage{relsize}
\usepackage[svgnames]{xcolor}

\usepackage{soul} 

\newcommand{\Uhll}{\mathbf{U}^{\rm hll}}
\newcommand{\Fhll}{\mathbf{F}^{\rm hll}}

\newcommand{\gammatm}{\gamma_{\mathsmaller{\mathrm{TM}}}}

\newcommand{\cross}[1]{}

\newcommand{\maa}[1]{\textcolor{Black}{#1}}

\newcommand{\tr}[1]{{\color{Black}{#1}}}

\newcommand{\CFL}{C_{\mathrm{CFL}}}

\title[An HLLC Riemann Solver for RRMHD]%
{An HLLC Riemann Solver for Resistive Relativistic Magnetohydrodynamics}

\author[S.~Miranda-Aranguren, M.\,A.~Aloy and T.~Rembiasz]%
{
  S.\,Miranda-Aranguren$^{1}$\thanks{sergio.miranda@uv.es},
  M.A.\,Aloy$^{1}$\thanks{miguel.a.aloy@uv.es}, 
  T.\,Rembiasz$^{1}$\\
$^{1}$Departamento de Astronom\'{\i}a y Astrof\'{\i}sica,  Universitat de Val\`encia,  C/. Dr.~Moliner 50, 46100 Burjassot, Spain  
}

\begin{document}

\date{Accepted XX 2018. Received XX 2017; in original form XX 2018}

\pagerange{\pageref{firstpage}--\pageref{lastpage}} \pubyear{2018}

\maketitle

\label{firstpage}

\begin{abstract}
  We present a new approximate Riemann solver for the augmented system
  of equations of resistive relativistic magnetohydrodynamics (RRMHD)
  that belongs to the family of Harten-Lax-van Leer contact wave
  (HLLC) solvers. In HLLC solvers,  the solution is approximated by two
  constant states flanked by two shocks separated by a contact
  wave. The accuracy of the new approximate solver is calibrated
  through one- and two-dimensional test problems.
\end{abstract}
%
%

%
\section{Introduction}
\label{sec:1}

Relativistic magnetised plasma is ubiquitously found among the most
violent and catastrophic phenomena of the Universe. Active galactic
nuclei (AGN) \citep{Blandford:2002}, Gamma-ray bursts \citep[GRBs;
][]{Lyutikov_Blandford:2003}, microquasars
\citep{Meier:2003,McKinney_Gammie:2004}, pulsars and magnetars
\citep{Bucciantini_etal:2005,Obergaulinger_Aloy:2017}, compact X-ray
binaries \citep{Varniere_etal:2002}, mergers of binary neutron stars
\citep{Rezzolla_etal:2011,Fernandez_Metzger:2016} or black holes
\citep{Marrone_etal:2007,Palenzuela_etal:2010,Vidal_etal:2015}, etc.,
may quite generically be endowed with dynamically relevant magnetic
fields.  From the dynamical point of view, magnetic fields play a main
role in the angular momentum transport required for driving accretion
in Keplerian discs girding compact objects.  The magnetorotational
instability \citep[MRI;][]{Velikhov:1959,Chandrasekhar:1960} is likely
to be the main mechanism inducing the angular momentum redistribution
in accretion discs
\citep[e.g.,][]{Balbus_Hawley:1991,Balbus_Hawley:1998}.  Another
context where the MRI seems to be crucial for the magnetic field to
reach dynamically relevant strength is stellar core collapse. The
post-collapsed core of a massive star develops suitable conditions for
the magnetic field to be amplified by the MRI
\citep[e.g.,][]{Akiyama_etal:2003,Obergaulinger_etal:2006,Cerda-Duran_etal:2007,Sawai_etal:2013,Moesta_etal:2015,Sawai_Yamada:2016}. In
any of these scenarios, the foremost question is not whether MRI may
develop, but instead, which is the primarily mechanism quenching the
magnetic field growth before it exhausts all the available free energy
(namely, differential rotational energy) in the system.
Compressibility caused by the magnetic field or parasitic
instabilities, which can be strongly affected or even are triggered by
non-ideal effects such as resistivity and viscosity
\citep{Goodman-Xu:1994,Latter_etal:2009,Pessah:2010}, are commonly
invoked as the main agents setting the termination level of the field
growth. In the last decade, the basic analytic models for magnetic
field saturation have been probed by means of numerical simulations in
different physical regimes including viscosity and resistivity
\citep[e.g.,][]{Rembiasz_etal:2016a,Rembiasz_etal:2016b,Guilet_Mueller:2015,Guilet_etal:2015},
though none of them has treated the problem of the MRI saturation
including relativistic effects.

Relativistic jets generated from compact objects are another example
where resistive effects may be instrumental to understanding their
generation. In one of the most accepted models of jet formation,
magnetic field taps a fraction of the rotational energy of a Kerr
black hole (BH) and launches a relativistic beam of very magnetised
plasma
\citep[e.g.][]{BZ77,Beskin_Kuznetsova:2000,Komissarov:2004,Okamoto:2006,McKinney:2006,Tchekhovskoy_etal:2010,Penna_etal:2013}. The
global structure of the magnetosphere surrounding either a neutron
star (NS) or a BH is reasonably well represented assuming that it is
force-free. However, it is unlikely that the force-free conditions
hold everywhere. For instance, in the case of NSs, there may exist
small regions (gaps) where particles are accelerated by the electric
field along the magnetic field lines. These procesesses may explain
the magnetospheric emission
\citep{Beloborodov_Thompson:2007,Beloborodov:2013a,Beloborodov:2013b,Levinson_Segev:2017}.
In the case of BH magnetospheres, it is not uncommon that even the
simplest topologies of the magnetic field encompass low latitude
regions where current-sheets \citep[which are known to be unstable
against the tearing mode (TM) instability;][]{Furth_etal:1963} are
present in the initially computed magnetospheric topologies
\citep[e.g., in split-monopole
configurations,][]{BZ77,Ghosh:2000,Contopoulos_etal:2013,Nathanail_Contopoulos:2014}. They
may also develop in the course of the dynamical evolution either
arising from the accretion disk
\citep{Goodman_Uzdensky:2008,Parfrey:2015} or due to MHD instabilities
of a Poynting-flux dominated flow
\citep{Eichler:1993,Begelman:1998,Giannios_Spruit:2006,Bromberg_Tchekhovskoy:2016}.

Beyond the undeniable dynamical influence that non-ideal resistive
effects may have on astrophysical sources associated to relativistic
outflows, the emission properties of such sources are likely bound to
the mechanisms of magnetic field dissipation if the outflows are
Poynting-flux dominated
\citep{Thompson:1994,Spruit_etal:2001,Giannios_Spruit:2005,Zhang_Yan:2011}. The
``internal shocks'' or ``shock-in-jet'' model
\citep[e.g.,][]{Rees:1978km,Rees:1994ca,Marscher:1985am,Spada:2001do,Bicknell:2002is,Mimica:2004ay}
has an important role in the understanding of the blazar spectra and
has been very often invoked for explaining many of the features of the
blazar variability and flares, as well as the prompt emission of
GRBs. For the latter, however, there have been claims that the
radiation efficiency is too low
\citep{Kumar:1999,Panaitescu_etal:1999,Kumar_Narayan:2009}, though
more detailed numerical models seem to ameliorate this potential
drawback \citep{Mimica_etal:2005}. Indeed, internal shocks happening
in a moderately magnetised plasma can be (much) more efficient
converting kinetic energy into thermal {\em and} magnetic energy than
purely hydrodynamic internal collisions
\citep{Mimica_etal:2007,Mimica_etal:2010}.  There are, however,
indications that the dissipation of magnetic fields may naturally
explain the observed phenomenology as we discuss in the next paragraphs.

First, employing particle-in-cell (PIC) simulations,
\cite{Sironi:2015} show that magnetic reconnection may deposit more
than $50\%$ of the dissipated energy into non-thermal leptons when the
magnetic field energy density is larger than the rest-mass energy
density. The shock downstream emitting region shows a rough
equipartition between magnetic field and radiating particles,
accounting naturally for this commonly observed property in
blazars. Along the same line, \cite{Petropoulou:2016} also conclude
that blazar flares naturally result from magnetic reconnection in a
magnetically dominated jet.

Second, the lack of a thermal component in the spectrum of some well
observed Fermi GRBs \citep[e.g., in GRB~080916C;][]{Zhang_Peer:2009}
is taken as an indication of the magnetisation of the radiating plasma
flow. The photospheric thermal component, which is expected to appear
in the standard ``fireball'' model, can be much dimmer (unobservable
in practice) if the outflow is Poynting dominated
\citep{Zhang_Meszaros:2002,Daigne_Mochkovitch:2002}, unless the
Poynting flux is directly converted into kinetic energy of the flow
below the photosphere \citep[e.g.,][]{Vlahakis_Koenigl:2003}. This
means that magnetic dissipation should play a fundamental role shaping the
observed spectra.

Third, the short time-scales displayed by the high-energy emission of
either GRBs or blazar jets in AGNs have driven the development of
various models where the reconnection of the magnetic field is of
paramount importance. Among them, we find the ``minijets'' or
``jets-in-a-jet'' model
\citep[e.g.,][]{Giannios_etal:2009,Giannios_etal:2010,Nalewajko_etal:2011,Giannios:2013,Barniol-Duran:2016}
and the ``fundamental emitters'' or ``relativistic turbulence'' model
\citep[e.g.,][]{Blandford:2002,Lyutikov_Blandford:2003,Lyutikov:2006,Kumar_Narayan:2009,Lazar_etal:2009,Narayan_Kumar:2009,Narayan_Piran:2012,ORiordan_etal:2017}.
The ``jets-in-a-jet'' model attributes the TeV emission in
relativistic AGN jets to blobs of plasma (plasmoids) where the
magnetic field dissipates by reconnection
\citep{Giannios_etal:2009,Giannios_etal:2010}. Under certain
conditions, the reconnection outflows are moderately relativistic and
may efficiently power the observed TeV flares through
synchrotron-self-Compton emission. The emission of the outflows
generated by episodic reconnection events is Doppler boosted by the
relativistic jet beam. The resulting TeV flares timescales may be as
short as minutes in the context of AGN jets, consistent with that
observed in, e.g., M87
\citep{Aharonian_etal:2006,Acciari_etal:2008,Albert_etal:2008} as 
well as in the blazars MrK 501 and PKS 2155-304
\citep{Aharonian_etal:2007,Albert_etal:2007}. We note, however, that
\cite{Narayan_Piran:2012} find that the emission properties
(variability) of relativistic turbulent motions in a beam with a
sufficiently fast jet (with a Lorentz factor $W >25$) accommodate more
easily the TeV light-curve of the blazar PKS 2155-304 than the
standard minijets model.

Fourth, models based upon the reconnection of the magnetic field have
also become popular to explain the prompt emission of GRBs. For
instance, the {\em Internal Collision-induced Magnetic Reconnection
  and Turbulence} (ICMART) model \citep{Zhang_Yan:2011,Deng_etal:2015}
assumes that in an intermittent magnetically dominated outflow
($1\lesssim \sigma_{\rm m} \lesssim 100$\footnote{The magnetisation is
  defined in Heaviside-Lorentz units as
  $\sigma_{\rm m}:=B^2/(\rho W^2c^2)$, where $B$, $\rho$, $W$, and $c$
  are the magnetic field strength, the rest-mass density, the bulk
  Lorentz factor of the plasma, and the speed of light in vacuum,
  respectively.}),
internal collisions take place. Due to the larger magnetisation of the
flow, the first generation of (weak) internal shocks happening at
relatively small distances from the central GRB engine
($\sim 10^{13}-10^{14}\,$cm) distort any large scale magnetic field
existing in the flow. The successive generations of shell collisions,
which take place at distance scales $\sim 10^{15}-10^{16}\,$cm,
trigger fast turbulent reconnection (an ICMART event) that may be
observed as a broad pulse in the GRB light curve.

Fifth, for typical long GRB jets, \cite{McKinney_Uzdensky:2012} find
that magnetic reconnection may be avoided due to the high collisional
rate of the plasma deep inside the GRB stellar progenitor, until a
``reconnection switch'' mechanism proceeds catastrophically near the
jet photosphere, where radiation is efficiently released.

Many relevant astrophysical phenomena involve shock waves. In the
specific context of this paper, the fast magnetic reconnection of
Petschek type \citep{Petschek:1964,Lyubarsky:2005} develops outflows
with Lorentz factors $W\sim \sigma_{\rm m}^{1/2}$. Indeed,
relativistic reconnection is an active area of research
\citep{Watanabe_Yokoyama:2006,Hesse_Zenitani:2007,Komissarov_etal:2007,Zenitani_Hoshino:2007,Zenitani_Hesse:2008,Zenitani_etal:2009,Tenbarge_etal:2010,Zenitani_etal:2010,Uzdensky:2011,Mizuno:2013,Takamoto:2013,Mohseni_etal:2015,DelZanna_etal:2016,Quian_etal:2017}. Reconnection
outflows are bounded by shocks that any numerical code aiming to study
them should handle well, in addition to current sheets and filaments
in the flow. In recent years\tr{,} a remarkable progress has been made
in numerical methods for resistive relativistic magnetohydrodynamics
\citep[RRMHD; e.g.,][]{Komissarov:2007,Palenzuela_etal:2009,Dumbser_Zanotti:2009,Takamoto_Inoue:2011,Bucciantini_DelZanna:2013,Takamoto:2014}.
\cross{Del Zanna et al (2016) performed the first simulations of the TM
instability in RRMHD.}  The implementation of such numerical methods is
chiefly based upon a conservative formulation of the RRMHD system of
equations. This requires evaluating either an exact or an approximate
solution to the Riemann problem at the interfaces between adjacent
computational zones (see, e.g., the excellent review by
\citealt{Marti_Mueller:2015}). Exact Riemann solvers usually are
computationally very expensive. Thus, approximate solvers have been
broadly used in classical magnetohydrodynamics (MHD) and in
relativistic MHD (RMHD) simulations. Among them, the Harten-Lax-van
Leer (HLL) solver \citep{HLL:1983} has been extensively used due to
its easy implementation and robustness.

Those properties of the HLL solver are based on its Jacobian-free
design, which avoids the decomposition of the jumps of the
characteristic variables over all right eigenvectors of the system of
equations. Instead, a single state that is an average of the solution
over the Riemann fan is computed. This single state is bounded by two
limiting waves where Rankine-Hugoniot (RH) jump conditions hold.  The
higher the number of different intermediate states the Riemann problem
develops (dictated by the number of different intermediate eigenvalues
of the Jacobian), the less accurately the single average state over
the Riemann wave structure represents the breakup of the discontinuity
(i.e., the more diffusive it is).  \maa{In its more basic form, the
  system of RRMHD equations constitutes a hyperbolic system of balance
  laws with two additional (elliptic) constraint equations that state
  the solenoidal character of the magnetic field and that the electric
  charge fixes the divergence of the electric field
  \citep{Dixon:1978,Anile:1989}. In the hyperbolic sector of the basic
  RRMHD system\tr{,} we find 12 eigenvalues corresponding to
  electromagnetic waves (6 eigenvalues), fast magnetosonic waves (2
  eigenvalues) and entropy, shear and charge waves (4
  eigenvalues). The preservation of the elliptic constrains in RRMHD
  can be enforced by a suitable constraint transport method
  \citep{Evans_Hawley:1988,Stone_Norman:1992} as in,
  e.g.~\cite{Bucciantini_DelZanna:2013}.}  In the formulation of RRMHD
of \cite{Komissarov:2007}, \maa{the basic RRMHD system is {\em
    augmented} with two additional equations, that control the
  evolution of both scalar potentials, which act as generalised
  Lagrangian multipliers \citep[GLM;][]{Dedner_etal:2002} to maintain
  the constraints of the electromagnetic field. In this formulation,
  the constraints are not elliptic equations but, instead, hyperbolic
  (telegrapher) equations. Thus, in the augmented system of RRMHD,}
there are 14 eigenvalues, 8 of which are degenerate and equal to the
speed of light (limiting the Riemann fan), in addition to two fast
magnetosonic waves and four contact waves moving at the local fluid
speed \citep{Cordero_etal:2012}. Therefore, the HLL single average
state spans 3 distinct intermediate states. In contrast, the system of
equations of RMHD has 5 different intermediate eigenvalues (two slow
magnetosonic, one Alfv\'en and one contact) under non-degenerate
conditions \citep{Anile:1989,Komissarov:1999,Anton_etal:2010}.

In order to ameliorate the deficiencies of the HLL-family of
approximate Riemann solvers, while at the same time keeping their
simplicity and computational efficiency, \cite{Toro_etal:1994}
proposed a generalisation of the HLL flux for the Euler
equations. These authors introduced an additional contact wave in the
solution separating two intermediate states and formulated a
Harten-Lax-van Leer contact (HLLC) wave approximate Riemann solver.
Since then, different HLLC Riemann solvers have been designed for MHD
\citep{Gurski:2004,Li:2005} and RMHD
\citep{Mignone_Bodo:2006,Honkkila_Janhunen:2007,Kim_Balsara:2014}.

In this paper, we present a new HLLC Riemann solver for the augmented
system of equations of RRMHD (Sec.\,\ref{sec:equations}). The extra
equations of the system include two scalar potentials, which control
the evolution of the solenoidal constraint on the magnetic field and
of the divergence of the electric field \citep{Komissarov:2007}. The
new solver is implemented in a generalisation of the MRGENESIS code
\citep{Aloy_etal:1999,Miranda_etal:2014}, which deals with the
stiffness of the RRMHD equations in the ideal limit employing either
Runge-Kutta Implicit-Explicit \citep[RK-IMEX;][]{Pareschi:2005} time
integrators or Minimally Implicit Runge-Kutta \citep[MIRK;
][]{Aloy_Cordero-Carrion:2016} methods.  The solver is obtained in
Sec.\,\ref{sec:HLLC1} for Resistive Special Relativistic MHD, but can
be used also in applications involving General Relativistic
gravitational fields resorting to the methodology devised in
\cite{Pons_etal:1998}. In order to demonstrate the numerical
capabilities of the new HLLC solver\tr{,} a number of standard
one-dimensional (1D) and two-dimensional (2D) numerical tests are
performed in Sec.\,\ref{sec:tests}.  In Sec.\,\ref{sec:TM}, we present
simulations of relativistic \tr{ideal} TMs.  We close this work with some
concluding remarks in Sec.\,\ref{sec:conclusions}.

\section{The RRMHD Equations}
\label{sec:equations}

A relativistic non-ideal magnetohydrodynamics fluid can be described by a system of balance laws \cite{Komissarov:2007}.
These equations express the conservation of charge, mass, momentum and energy, together with the Maxwell equations.  In
the Heaviside-Lorentz units used in this paper, with the speed of light $c =1$, these equations read:
\begin{align}
 \partial_t q          &= - \nabla \cdot \mathbf{J},           \label{eq:chargecons}\\ 
 \partial_t \psi       &=  -   \nabla \cdot \mathbf{E} \ \ + q - \kappa \psi,    \label{eq:psi}    \\
 \partial_t \phi       &=  -   \nabla \cdot \mathbf{B} \ \ - \kappa \phi,         \label{eq:phi}   \\  
 \partial_t \mathbf{E} &=  \ \ \nabla \times \mathbf{B} - \nabla \psi - \mathbf{J},  \label{eq:E-Maxwell}\\  
 \partial_t \mathbf{B} &=  -   \nabla \times \mathbf{E} -    \nabla \phi, \label{eq:B-Maxwell}\\
 \partial_t D          &= - \nabla \cdot \mathbf{F}_D,         \label{eq:masscons}\\ 
\partial_t \mathcal{E}    &= - \nabla \cdot \mathbf{F}_{\tau},    \label{eq:energycons} \\
  \partial_t \mathbf{S} &= - \nabla \cdot \mathbf{F}_{\mathbf{S}}, \label{eq:momentumcons}
\end{align} 
where $q$, $\mathbf{E}= (E_x,E_y,E_z)^T$, $\mathbf{B}= (B_x,B_y,B_z)^T$,
$D$, $\mathcal{E}$ and $\mathbf{S}$ stand for the charge density, the
electric and magnetic field 3-vectors in the laboratory frame, the
relativistic mass-density, the relativistic energy density and the
momentum density, respectively. We use a divergence cleaning strategy
\citep{Dedner_etal:2002}, where the scalar pseudopotentials
$\psi$ and $\phi$ enforce the conservation of $q$ and of the
solenoidal constrain $\nabla \cdot \mathbf{B}=0$,
  respectively. These pseudopotentials decay exponentially with time
if the constant $\kappa:= c_h^2/c_p^2> 0$, where $c_p^2$ can be
regarded as a diffusion coefficient and $c_h$ as the finite speed at
which $\nabla \cdot \mathbf{B}$ errors propagate. For simplicity and
not to limit further the time step, we choose this finite speed equal
to speed of light ($c_h = 1$) and following
\cite{Mignone_Tzeferacos:2010}, we define the dimensionless parameter,
\begin{equation*}
\alpha := \Delta h \frac{c_h}{c^2_p},
\end{equation*}
where $\Delta h := \min(\Delta x, \Delta y, \Delta z)$, and $\Delta x$, $\Delta y$, $\Delta z$ are the grid spacings in
the three Cartesian directions. According to \cite{Mignone_Tzeferacos:2010}, the errors associated with the violation of
the magnetic field solenoidal constrain are minimised when $\alpha \in [0,1]$. In the numerical experiments presented in
this work (Sec.\,\ref{sec:tests}), we set $\alpha =1$.

The {\em augmented} system of RRMHD equations (\ref{eq:chargecons})$-$(\ref{eq:momentumcons}), can be written as a
system of balance laws:
\begin{equation}
\partial_t \mathbf{U} + \sum_{k} \partial_k \mathbf{F}(\mathbf{U}) = \mathbf{\Omega}(\mathbf{U}),\label{eq:RRMHDsystem}
\end{equation}
where $\mathbf{U}$, $\mathbf{F}$ and $\mathbf{\Omega}$ are the
conserved variables, the fluxes and the source terms,
respectively. The vector
$\mathbf{U} = (\psi, \phi, \mathbf{E}, \mathbf{B}, q, D, {\cal E},
\mathbf{S})^T$ of conserved variables is related to the vector of
primitive or physical variables
$\mathbf{W}= (\psi, \phi, \mathbf{E}, \mathbf{B}, q, \rho, p_{\rm g},
\mathbf{v})^T$ through the following algebraic relations:%
\footnote{The subset of values $\{\phi,\psi,\mathbf{E},\mathbf{B},q\}$ can
  be regarded as both conserved or primitive variables.}
\begin{align}
D            &= \rho W,  \\
{\cal E}       &= {\cal E}_{\rm EM} + {\cal E}_{\rm hyd} =\frac{1}{2} (\mathbf{E}^2 + \mathbf{B}^2) + \rho h W^2 - p_{\rm g}, \label{eq:calE}   \\
\mathbf{S} &= \mathbf{S}_{\rm EM} + \mathbf{S}_{\rm hyd}=\mathbf{E}
             \times \mathbf{B} + \rho h W^2 \mathbf{v}, \label{eq:momentumdensity}
\end{align}
where $p_{\rm g}$ is the gas pressure, $\rho$ the proper rest mass
density, $\mathbf{v} = (v_x, v_y, v_z)^T$ the fluid velocity measured in
laboratory  frame, $W = (1 - \mathbf{v}^2)^{-1/2}$ the Lorentz factor and
$h=1+\varepsilon+p_{\rm g}/\rho$ the specific enthalpy, $\varepsilon$
being the specific energy. For an ideal gas equation of state with
constant adiabatic index $\gamma$, $p=(\gamma-1)\rho\varepsilon$, the
specific enthalpy can be written as
 \begin{equation*}
\maa{
 h(\rho,p) = 1 + \frac{\gamma}{\gamma -1} \frac{p}{\rho} ,}
\end{equation*}
and the local sound speed reads $c_{\rm s} = \sqrt{\gamma p / (\rho
  h)}$. The system (\ref{eq:RRMHDsystem}) is hyperbolic if
$\partial_\varepsilon p_{\rm g}-\rho c_{\rm s}^2\ne 0$ \citep{Cordero_etal:2012}.

The fluxes can be explicitly written in terms of the primitive variables:
\begin{equation}
  \begin{aligned}
\mathbf{F}_q &= \mathbf{J} ,\\
\mathbf{F}_\psi &= \mathbf{E} ,\\
\mathbf{F}_\phi &= \mathbf{B} ,\\
F_{E^i}^j &= \delta^j_i \psi -\epsilon_{ijk}B^k ,\\
F_{B^i}^j &= \delta^j_i \phi +\epsilon_{ijk}E^k ,\\
\mathbf{F}_D &= \rho W \mathbf{v} ,\\
\mathbf{F}_{{\cal E}}  &=  \mathbf{E} \times \mathbf{B} + \rho h W^2 \mathbf{v}  ,\\
\mathbf{F}_{\mathbf{S}}  &= - \mathbf{E} \mathbf{E}  -
                                          \mathbf{B} \mathbf{B}  +
                                          \rho h W^2 \mathbf{v}\mathbf{v}+ P\boldsymbol{g}  ,
\end{aligned}
\label{eq:Fluxes}
\end{equation}
where $P= p_{\rm g} + (\mathbf{E}^2 + \mathbf{B}^2) /2$ is the total
pressure, $\boldsymbol{g} = \text{diag}(-1,1,1,1)$ is the Minkowski
metric tensor and $\delta^j_i$ and $\epsilon_{ijk}$ are the Kronecker
delta and the Levi-Civita symbol, respectively. The source term is
given by
$\mathbf{\Omega} = (q - \kappa \psi , - \kappa \phi , - \mathbf{J} ,
\mathbf{0}_{1X9})^T$, where $\mathbf{J} = (J_x, J_y, J_z)^T$ is the
electric current density, which employing the Ohm's law takes the form
\citep{Komissarov:2007}
\begin{equation}
  \mathbf{J} := \mathbf{J}_{\rm s} + q \mathbf{v} = \sigma W [\mathbf{E} + \mathbf{v} \times \mathbf{B} - (\mathbf{E} \cdot \mathbf{v})\mathbf{v} ] + q \mathbf{v},
\end{equation}
where $\sigma$ is the electrical conductivity of the medium. In the ideal
case (i.e., $\sigma \to \infty$) the resistive part ($\mathbf{J}_{\rm s}$)
of the total electric current density makes the system of RRMHD
equations stiff. In that limit, the characteristic time scale
($1/\sigma$) of the source term ($\Omega$) is, in general, much
shorter than the timescale of hyperbolic part ($\mathbf{F}$). This may
introduce instabilities in the time integration if an explicit
method is directly used. We face these possibility employing
implicit-explicit methods. In particular, we may use either RK-IMEX
\citep{Palenzuela_etal:2009} or MIRK 
\citep{Aloy_Cordero-Carrion:2016} schemes. In the case of using an RK-IMEX scheme,
the source vector $\mathbf{\Omega}$ is split into both a
stiff and a non-stiff operators, which read,
$\mathbf{\Omega}_{\rm s} = (0,0 , - \mathbf{J}_{\rm
  s},\mathbf{0}_{1X9})$ and
$\mathbf{\Omega}_{\rm ns} = (q - \kappa \psi , - \kappa \phi , -
q\mathbf{v} , \mathbf{0}_{1X9})$, respectively.

Note that in Eqs.\,(\ref{eq:calE}) and (\ref{eq:momentumdensity}), we explicitly write the total conserved energy
density and momentum density as the sum of two independent contributions, electromagnetic (subscript ``EM'') and purely
hydrodynamic (subscript ``hyd'').  In the case of using a MIRK time integration scheme, the recovery of the primitive
variables from the conserved ones reduces to the same algorithm as for the relativistic Euler equations
\citep[][App.\,C]{Aloy_etal:1999}, taking as conserved variables the subset $\{D,\mathbf{S}_{\rm hyd},\tau_{\rm
  hyd}\}=\{D, \mathbf{S}-\mathbf{S}_{\rm EM},{\cal E}-{\cal E}_{\rm  EM}-D\}$.  This is because , firstly, the subset of
variables $\{\phi,\psi,\mathbf{E},\mathbf{B},q\}$ are both conserved and primitive variables and, secondly, the RRMHD
system includes both equations for $\mathbf{E}$ (Eq.\,\ref{eq:E-Maxwell}) and $\mathbf{B}$ (Eq.\,\ref{eq:B-Maxwell}).
We note, however, that for generic RK-IMEX time integration schemes, the simplest version of the procedure to recover
variables cannot be directly used. Instead, we resort to the procedure delineated in Sec.\,4.2 of
\cite{Palenzuela_etal:2009}.

\section{Characteristic speeds}
\label{sec:characteristic}

We consider for simplicity a Riemann problem along the
$x-$coordinate direction set at the location $x_{i+1/2}$, i.e., the
interface between two consecutive cells, $i$ and $i+1$
\begin{equation*}
  \begin{split}
  & \mathbf{U}(x,0) = \left\{
  \begin{aligned}
    \mathbf{U}_{l}  & \quad \text{if} \  x < x_{i+1/2} \\
    \mathbf{U}_{r}  & \quad \text{if} \  x > x_{i+1/2},
  \end{aligned} \right.
  \end{split}
\end{equation*}
where $\mathbf{U}_{l}$ and $\mathbf{U}_{r}$ are the initial (uniform)
states to the left and to the right of $x_{i+1/2}$.  The breakup of
this Riemann problem in RRMHD produces a set of 14 waves with
characteristic speeds
\begin{align}
 \hspace{-0.25cm}&\lambda_{EB_\pm} =\pm 1, & \text{(multiplicity 4)} \label{eq:lambdaEB}\\
 \hspace{-0.25cm}&\lambda_q          =v_x, & \text{(multiplicity 1)}\label{eq:lambdaq}\\
 \hspace{-0.25cm}&\lambda_{H_0}    =v_x, & \text{(multiplicity 3)}\label{eq:lambdaH0}\\
 \hspace{-0.25cm}&\lambda_{H_\pm} =v_x\Delta_H \pm c_{\rm s}\sqrt{\frac{\rho h}{\Delta}(1-v_x^2\Delta_H)},  \hspace{-0.16cm}& \text{(multiplicity 1)}
\label{eq:lambdaHpm}
\end{align}
 where
$\Delta:=\rho h W^2(1-\mathbf{v}^2c_{\rm s}^2)$, and
$\Delta_H:=(1-c_{\rm s}^2)/(1-\mathbf{v}^2c_{\rm s}^2)$
\citep{Cordero_etal:2012}.

\maa{We point out that the characteristic speeds listed in
  Eqs.\,(\ref{eq:lambdaEB})-(\ref{eq:lambdaHpm}) are computed,
  following the standard practice for hyperbolic systems of partial
  differential equations \cite[e.g.,][]{Anile:1989} as the eigenvalues
  of the Jacobian matrices of the system of RRMHD equations. These
  eigenvalues represent the wave speeds of plasma perturbations only
  in the infinite resistivity limit. This is the reason why Alfv\'en
  and slow magnetosonic waves do not explicitly appear among the
  obtained characteristic speeds. The presence of source terms (the
  current in Ampere's law; Eq.\,(\ref{eq:E-Maxwell}) and the GLM
  scalar potentials in Eqs.\,(\ref{eq:psi}) and (\ref{eq:phi})) alters
  these signal velocities. Indeed, both the basic and the augmented
  RRMHD systems belong to the class of ``hyperbolic systems of
  conservation laws with relaxation'' as defined by, e.g.,
  \cite{Whitham:1974}.  The latter term denotes hyperbolic systems of
  $n$ partial differential equations in conservation form with source
  terms, which have as a limit a hyperbolic system of $M$ ($M<n$)
  equations called the equilibrium system as $n-M$ relaxation time
  parameters $\tau_i\rightarrow 0$. In our case, the equilibrium
  system is the one formed by the equations of (ideal) RMHD and the
  relaxation parameters are the resistivity, $\eta:=1/\sigma$, and
  $1/\kappa$. Following the convention of \cite{Pember:1993}, the
  characteristic speeds of the equilibrium and non-equilibrium systems
  are called {\em equilibrium} and {\em frozen} characteristic speeds,
  respectively.\footnote{\maa{One can actually define the concept of
      stiffness in this framework saying that a system of conservation
      laws is stiff when at least one of its relaxation times is small
      compared to the time scale determined by the frozen
      characteristic speeds of the system and some appropriate length
      scale.}} }

\maa{The equilibrium ($\tilde{\lambda}_j$, $j=1,\ldots,M$) and frozen
  ($\lambda_j$, $j=1,\ldots,n$) characteristic speeds satisfy the {\em
    subcharacteristic condition} if they are interlaced, i.e., if each
  $\tilde{\lambda}_j$ lies in the closed interval
  $[\lambda_k,\lambda_{k+n-M}]$.  The previous condition is necessary
  for the stability of linearised systems with relaxation
  \citep{Whitham:1974}. The subcharacteristic condition is satisfied
  by the characteristic speeds of both the basic and the augmented
  RRMHD systems and the characteristic speeds of the (ideal) RMHD
  system. Therefore, the numerical solution of the former systems
  tends to the solution of the equilibrium (ideal RMHD) system as the
  relaxation time tends to zero \citep{Chen_etal:1994}.  The
  eigenspeeds computed by \cite{Cordero_etal:2012} only apply to the
  frozen limit where electromagnetic phenomena and matter are
  completely decoupled. However, the coupling is restored by our
  method of lines and the application of partly implicit time
  integration methods (either RKIMEX or
  MIRK). \cite{Takamoto_Inoue:2011} took an alternative approach
  employing a method of characteristics (MOC) for the integration of
  the RRMHD equations. Alfv\'en modes are explicitly restored in their
  approach \cite{Takamoto_Inoue:2011} by modifying Ampere's law
  introducing an effective propagation speed, which depends on the
  damping rate of electromagnetic modes.  That is, when all
  electromagnetic modes (moving at the speed of light) are damped
  during one time step of the numerical evolution, $\Delta t$, they
  use appropriate (ideal RMHD) characteristic speeds for their MOC;
  otherwise, they resort to using the speed of light.  In practice,
  the method of \cite{Takamoto_Inoue:2011} only uses characteristic
  speeds smaller than the speed of light when
  $\Delta t < \Delta t_{\rm e}:= 4\pi/\sigma$. Our partly implicit
  time integration methods typically provide values of
  $\Delta t = \CFL \Delta x \gg \Delta t_{\rm e}$ for reasonable
  values of the grid spacing. This means that our algorithm should
  yield a qualitatively similar restoration of the Alfv\'en, fast and
  slow RMHD modes as that of \cite{Takamoto_Inoue:2011} in practical
  applications.}

%

\section{HLLC Solver}
\label{sec:HLLC1}

An HLLC approximate Riemann solver avoids the full characteristic decomposition of all the wave pattern in the Riemann
fan by only introducing a contact wave with constant speed, $\lambda^{*}$, which separates two intermediate states
($\mathbf{U}_l^*, \mathbf{U}_r^*$) bounded by two fast shocks. More explicitly, the solution of the initial value
problem in each cell interface is written as
\begin{equation}
  \begin{split}
  & \mathbf{U}(0,t) = \left\{
  \begin{aligned}
    \mathbf{U}_{l}       & \quad \text{if} \  \lambda_l \ge 0 \\
    \mathbf{U}_{l}^*     & \quad \text{if} \  \lambda_l < 0 \le \lambda^* \\
    \mathbf{U}_{r}^*     & \quad \text{if} \  \lambda^* < 0 \le \lambda_r \\
    \mathbf{U}_{r}       & \quad \text{if} \  \lambda_r <  0,
  \end{aligned} \right.
  \end{split}
\label{eq:Uhllc}
\end{equation}
where $\lambda^*$ is the propagation velocity of the contact wave
and $\lambda_l,\lambda_r$ are estimates of the maximum signal speeds
propagating to the left and to the right of the initial discontinuity,
respectively. Since in RRMHD the electromagnetic fields propagate at
the speed of light (Eq.\,\ref{eq:lambdaEB}), we set $\lambda_l =-1$
and $\lambda_r=1$, but for sake of clarity, we maintain the notation
$\lambda_l$, $\lambda_r$ throughout the paper.

The numerical fluxes corresponding to the assumed solution
(\ref{eq:Uhllc}) are:
\begin{equation}
  \begin{split}
  & \Tilde{\mathbf{F}} = \left\{
  \begin{aligned}
    \mathbf{F}_{l}       & \quad \text{if} \  \lambda_l \ge 0 \\
    \mathbf{F}_{l}^*     & \quad \text{if} \  \lambda_l < 0 \le \lambda^* \\
    \mathbf{F}_{r}^*     & \quad \text{if} \  \lambda^* < 0 \le \lambda_r \\
    \mathbf{F}_{r}       & \quad \text{if} \  \lambda_r < 0,
  \end{aligned} \right.
  \end{split}
\label{eq:tildeF}
\end{equation}
with $\mathbf{F}_{l}=\mathbf{F}(\mathbf{U}_l)$ and
$\mathbf{F}_{r}=\mathbf{F}(\mathbf{U}_r)$ and $\mathbf{F}_{l}^*$,
$\mathbf{F}_{r}^*$ the intermediate flux functions, whose expressions
will be found in the next subsections.

\subsection{Rankine-Hugoniot conditions for a bounded source term}
Following \cite{LeVeque:2002}, a one-dimensional, scalar balance law, $\partial_t U(x,t) + \partial_x F(U(x,t))
= \Omega(x,t)$, can be expressed in  integral form as:
\begin{align}
  \int_{x_{1}}^{x_{1} + \Delta x}  U(x,t_1 + \Delta t) \ dx -
  \int_{x_{1}}^{x_{1} + \Delta x}  U(x,t_1 ) \ dx = &\notag \\
  \int_{t_1}^{t_1+\Delta t} F(U(x_1,t) dt - \int_{t_1}^{t_1+\Delta t} F(U(x_1+ \Delta x,t) dt  &\notag \\
                                                       +   \int_{t_1}^{t_1+\Delta t}  \int_{x_{1}}^{x_{1} + \Delta x} \Omega(x,t) \ dx \ dt, &\label{eq:hllc_000}
\end{align}
where the integration domain is
$[t_1,t_1+\Delta t]\times [x_1,x_1+\Delta x]$. If the
source term $\Omega(x,t)$ is a bounded function, the last integral in
(\ref{eq:hllc_000}) can be expressed in terms of the characteristic
velocity of the balance law, $\lambda$, as,
\begin{align}
  \int_{t_1}^{t_1+\Delta t}  \int_{x_{1}}^{x_{1} + \Delta x} \Omega(x,t) \ dx \ dt &\approx \Delta t \ \Delta x \ \Omega(x,t) \notag \\
                                                                        &\approx \Delta t^2 \ \lambda \ \Omega(x,t) \label{hllc_000b}.
\end{align}
Thus, from the conservation law defined in (\ref{eq:hllc_000}) the RH
conditions in vectorial form, can be written as:
\begin{equation*}
\lambda (\mathbf{U}_{r} - \mathbf{U}_{l}) \approx \mathbf{F}_{r} - \mathbf{F}_{l} + \lambda \ \Delta t \ \mathbf{\Omega},
\end{equation*}
which is an expression of the order $\mathcal{O}(\Delta t)$ and the contribution of the source term vanishes as $\Delta
t \to 0$. Hence, a bounded source term does not change the RH conditions at a discontinuity and, thus, we can use the
following set of RH conditions for the shocks and contact wave in the HLLC solution:
\begin{align}
  \lambda_l (\mathbf{U}_{l}^* - \mathbf{U}_{l}  ) &= \mathbf{F}_{l}^* - \mathbf{F}_{l} , \label{eq:RHl}\\
  \lambda^* (\mathbf{U}_{r}^* - \mathbf{U}_{l}^*) &= \mathbf{F}_{r}^* - \mathbf{F}_{l}^* , \label{eq:RH} \\
  \lambda_r (\mathbf{U}_{r}   - \mathbf{U}_{r}^*) &= \mathbf{F}_{r}   - \mathbf{F}_{r}^*. \label{eq:RHr} 
\end{align}
\subsection{HLLC consistency conditions}
\label{sec:HLLCconsistency}
Following \cite{Mignone_Bodo:2006}, we find the consistency conditions for
conserved intermediate variables by adding the tree equations in \maa{(\ref{eq:RHl})-(\ref{eq:RHr})},
\begin{equation}
\begin{aligned}
\frac{\left(\lambda^* - \lambda_l \right) \mathbf{U}_{l}^* + \left(\lambda_r - \lambda^* \right) \mathbf{U}_{r}^*}{\lambda_r - \lambda_l } &= \Uhll,
\end{aligned}\label{eq:Uhll}
\end{equation}
where
\begin{equation*}
\begin{aligned}
\Uhll &= \frac{\lambda_r \mathbf{U}_{r} - \lambda_l \mathbf{U}_{l} + \mathbf{F}_{l} - \mathbf{F}_{r}}{\lambda_r - \lambda_l },
\end{aligned}\label{hllc_003}
\end{equation*}
is the (HLL) integral average of the solution of the Riemann problem
over the wave fan (cf. \citealt{Toro:1997}, \S\,10.4). Likewise,
dividing each equation in \maa{(\ref{eq:RHl})-(\ref{eq:RHr})} by their corresponding
$\lambda$ on the left-hand sides and adding all the resulting
expressions one finds 
\begin{equation}
\begin{aligned}
\frac{\left(\lambda^* - \lambda_l \right) \lambda_r \mathbf{F}_{l}^* +  \left(\lambda_r - \lambda^* \right) \lambda_l  \mathbf{F}_{r}^*}{\lambda_r - \lambda_l } &= 
\lambda^* \Fhll
\end{aligned}
\label{eq:lambda*Fhll}
\end{equation}
with
\begin{equation}
\Fhll = \frac{\lambda_r \mathbf{F}_{l} - \lambda_l \mathbf{F}_{r} + \lambda_r \lambda_l (\mathbf{U}_{r} - \mathbf{U}_{l})}{\lambda_r - \lambda_l }
\label{eq:Fhll}
\end{equation}
being the (HLL) flux integral average of the solution over the Riemann
wave structure.  We point out that employing the (global) values for
the system limiting speeds $\lambda_l=-1$ and $\lambda_r=+1$, the HLL
flux (\ref{eq:Fhll}) reduces to the Local Lax-Friedrich flux.

Building an HLLC solver for the RRMHD system of equations ultimately
amounts to obtaining the expressions of $\mathbf{F}^*_l$ and
$\mathbf{F}^*_r$ in (\ref{eq:tildeF}). In general,
$\mathbf{F}^*_l\ne\mathbf{F}(\mathbf{U}^*_l)$ and
$\mathbf{F}^*_r\ne\mathbf{F}(\mathbf{U}^*_r)$. Instead, the RH jump
conditions (\ref{eq:RH}) provide a set of relations between the
starred numerical fluxes and the corresponding state vectors
$\mathbf{U}^*_l$ and $\mathbf{U}^*_r$. More precisely, if
$\mathbf{U}^*_l$ and $\mathbf{U}^*_r$ are known, from
\maa{Eqs.\,(\ref{eq:RHl})} and \maa{(\ref{eq:RHr})}, we have
\begin{eqnarray}
  \mathbf{F}_{l}^*  &=& \mathbf{F}_{l}  + \lambda_l (\mathbf{U}_{l}^* - \mathbf{U}_{l}  ) , \notag\\
  \mathbf{F}_{r}^*  &=& \mathbf{F}_{r}  + \lambda_r (\mathbf{U}_{r}^* - \mathbf{U}_{r}  ). \label{eq:RH2}
\end{eqnarray}

If each state vector has $n$ components ($n=14$ in the RRMHD system of
equations (\ref{eq:chargecons})$-$(\ref{eq:momentumcons})), the
problem at hand has $4n+\maa{1}$ unknowns,
i.e.~$\mathbf{F}^*_l, \mathbf{F}^*_r, \mathbf{U}^*_l, \mathbf{U}^*_r$,
and $\lambda^*$. \maa{In addition, there are four ancillary variables,
  $v^*_{x,r}$, $v^*_{x,l}$, $P^*_r$ and $P^*_l$, originating from the
  fact that the fluxes (Eq.\,\ref{eq:Fluxes}) cannot be expressed in
  closed form only as a function of the conserved variables.  However,
  if we include these ancillary variables as a part of our problem,
  the momentum in the $x-$direction is not an independent variable,
  since using Eqs.\,(\ref{eq:calE}) and (\ref{eq:momentumdensity}), it
  can be written as}
\begin{equation}
S_{x,a}^* = (\mathbf{E}_a^* \times \mathbf{B}_a^*)_x + ({\cal E}_a^* + P_a^* -(\mathbf{E}_a^*)^2 - (\mathbf{B}_a^*)^2) v_{x,a}^*,
\label{eq:momentumdensity*}
\end{equation}
\maa{where the subscript $a=l,r$ refers to each of the two states to
  the left and to the right of the contact wave.  Thus, we have to
  find $4n+5$ relations among the unknowns. Two of them are provided
  by Eq.\,(\ref{eq:momentumdensity*}). There are $2n$ consistency
  relations (\ref{eq:Uhll}) and (\ref{eq:lambda*Fhll}).  We advance
  that part of the $n$ RH conditions across the middle wave
  (Eq.\,\ref{eq:RH}) become trivially satisfied (and thus replaced) by
  the subsequent assumptions that we employ to find the remaining
  $2n+3 (=31)$ equations. We chose these additional equations in the
  following way. First, we assume the continuity of a subset of the
  conserved variables and of the four ancillary variables. Second, we
  assume the functional dependence of the fluxes in the starred
  regions with the conserved and ancillary variables. Specifically, we
  impose continuity of the charge density, of the scalar potentials
  $\phi$ and $\psi$ as well as of the magnetic and electric field
  components across the contact wave,
\begin{equation}
\begin{split}
\begin{aligned}
 \hspace{-.3cm} q^*_r  &= q^*_l :=q^*,  \hspace{-.1cm}& B^*_{i,r}  &= B^*_{i,l} :=B^*_i, \hspace{-.1cm}& E^*_{i,r} &= E^*_{i,l} :=E^*_i ,\\
 \hspace{-.3cm}\phi^*_r  &= \phi^*_l :=\phi^* , \hspace{-.1cm}&  \psi^*_r  &= \psi^*_l :=\psi^*, \hspace{-.1cm}&  \label{eq:continuity_ContactWave}
\end{aligned} 
\end{split}
\end{equation}
where $i=x,y,z$. For the ancillary variables, we impose continuity of
the total pressure and \tr{of} the normal component of the velocity, i.e.,
\begin{equation}
\begin{split}
\begin{aligned}
 P^*_r  &= P^*_l :=P^* , \: &  v^*_{x,r}  &= v^*_{x,l} =\lambda^*.
\end{aligned} 
\end{split}
\label{eq:continuity_ContactWave1}
\end{equation}
}

\maa{The conditions imposed on the conserved variables
  (Eq.\,\ref{eq:continuity_ContactWave}) provide 9 relations among the
  unknowns of our problem, to which we add the 9 non-trivial RH
  conditions across the middle wave corresponding to the same subset
  of conserved variables (i.e., $q^*$, $\phi^*$, $\psi^*$,
  $\mathbf{E}^*$ and $\mathbf{B}^*$), as well as the additional three
  conditions (Eq.\,\ref{eq:continuity_ContactWave1}) on the ancillary
  variables. Therefore, there are still 10 additional missing
  equations to set a well posed problem, which come from assuming that
  the form of the fluxes $\mathbf{F}^*$ formally is the same as that
  of Eq.\,(\ref{eq:Fluxes}) in the $x$-direction for the following set
  of variables $(D^*,S_x^*,S_y^*,S_z^*,{\cal E}^*)$. More precisely,
  we assume that the starred fluxes relate to the starred variables
  through
\begin{align}
\hspace{-0.3cm}F^*_{D,a} &= D_a^* \lambda^* \label{eq:Flux*D},\\
\hspace{-0.3cm}F^*_{{\cal E},a}  &=  S^*_{x,a} \label{eq:Flux*calE},\\
\hspace{-0.3cm}F^*_{S_x,a}  &= - E^*_x E^*_x - B^*_x B^*_x + [S^*_{x,a} - (\mathbf{E}^* \times \mathbf{B}^*)_x]\lambda^* + P^*\label{eq:Flux*Sx},\\
\hspace{-0.3cm}F^*_{S_y,a}  &= - E^*_x E^*_y - B^*_x B^*_y + [S^*_{y,a} -  (\mathbf{E}^* \times \mathbf{B}^*)_y]\lambda^*\label{eq:Flux*Sy},\\
\hspace{-0.3cm}F^*_{S_z,a}  &= - E^*_x E^*_z - B^*_x B^*_z + [S^*_{z,a} -  (\mathbf{E}^*\times \mathbf{B}^*)_z]\lambda^*.\label{eq:Flux*Sz}
\end{align}
The assumed dependence of the fluxes on the conserved variables and on
the $\lambda^*$ and $P^*$ makes that 5 RH conditions across the
contact wave (Eq.\,\ref{eq:RH}) corresponding to the subset of
variables $(D^*,S_x^*,S_y^*,S_z^*,{\cal E}^*)$ become trivially
satisfied. Hence, these equations are replaced by
Eqs.\,(\ref{eq:Flux*D})-(\ref{eq:Flux*Sz}).  }

For the subset of conserved variables for which we have assumed continuity across
the contact wave (Eq.\,\ref{eq:continuity_ContactWave}), the
consistency condition (\ref{eq:Uhll}) yields:
\maa{
\begin{equation}
\begin{split}
\begin{aligned}
 \hspace{-.3cm} q^*  &=q^{\rm hll},  \hspace{-.1cm}& B^*_{i}  &=B^{\rm hll}_i, \hspace{-.1cm}& E^*_i &= E^{\rm hll}_i ,\\
 \hspace{-.3cm}\phi^*  &= \phi^{\rm hll} , \hspace{-.1cm}&  \psi^*  &= \psi^{\rm hll}. \hspace{-.1cm}& 
\end{aligned} 
\end{split}
\label{eq:continuity_ContactWave2}
\end{equation}
}
The fluxes corresponding to the conserved variables given in
(\ref{eq:continuity_ContactWave2}) are readily found applying
Eqs.\,(\ref{eq:RH2}):
\maa{
\begin{equation}
\begin{aligned}
F^*_{q,a}  &=F^{\rm hll}_q ,  & F^*_{\phi,a}  &=F^{\rm hll}_\phi ,  & F^*_{\psi,a}  &=F^{\rm hll}_\psi ,  \\
F^*_{B_i,a}  & =F^{\rm hll}_{B_i}, & F^*_{E_i,a} & =F^{\rm hll}_{E_i}.
\end{aligned}
\label{eq:Fluxes*q-B-E}
\end{equation}
}


%
\maa{The flux of ${\cal E}^*_a$ is exactly $S^*_{x,a}$
  (Eq.\,\ref{eq:Flux*calE}) and, hence, employing the consistency
  relations (\ref{eq:lambda*Fhll}) and (\ref{eq:Uhll}), we find
\begin{equation}
\begin{aligned}
S_{x,a}^* &= \frac{ \lambda_a S_{x}^{hll} - \lambda^* F_{\cal E}^{hll}}{\lambda_a-\lambda^*}. 
\end{aligned}
\label{eq:Ua*_1}
\end{equation}
}


\maa{Similarly, the especially simple form of the flux
  of $D$ (Eq.\,\ref{eq:Flux*D}), inserted in any of the Eqs.\,(\ref{eq:RH2}),
  yields:
\begin{equation}
\begin{aligned}
D_a^*   &=  \frac{\lambda_a - v_{x,a}}{\lambda_a - \lambda^*}D_a .  
\end{aligned}
\label{eq:D*2}
\end{equation}
}

Using relation (\ref{eq:lambda*Fhll}) for the $y-$ and $z-$momentum
density fluxes (Eqs.\,\ref{eq:Flux*Sy} and \ref{eq:Flux*Sz}) combined
with Eq.\,(\ref{eq:Uhll}) for $S^{hll}_y$ and $S^{hll}_z$ yields
\begin{equation}
\begin{aligned}
S_{y,a}^* &= \frac{ \lambda_a S_{y}^{hll} - F_{S^y}^{hll} - \lambda^*  (\mathbf{E}^* \times \mathbf{B}^*)_y - E^*_x E^*_y - B^*_x B^*_y}{\lambda_a-\lambda^*}, \\
S_{z,a}^* &= \frac{ \lambda_a S_{z}^{hll} - F_{S^z}^{hll} - \lambda^*  (\mathbf{E}^* \times \mathbf{B}^*)_z - E^*_x E^*_z - B^*_x B^*_z}{\lambda_a-\lambda^*}. \\
\end{aligned}
\label{eq:S*_yz}
\end{equation}

Applying the consistency relation (\ref{eq:Uhll}) to the  $x$
component of the momentum density, we find
\begin{equation} 
S_x^{hll}  =  \lambda^* [{\cal E}^{hll} + P^* -  (\mathbf{E}^{*})^2 -
(\mathbf{B}^{*})^2]  + (\mathbf{E}^* \times \mathbf{B}^*)_x.
\label{eq:Shll}
\end{equation}
Likewise, applying the consistency relation (\ref{eq:lambda*Fhll}) to
the $x$  component of the momentum density flux
(\ref{eq:Flux*Sx}) and to the energy density flux (\ref{eq:Flux*calE}) yields
\begin{equation}
\lambda^* F^{hll}_{S^x} =  (\lambda^*)^2 [ F^{hll}_{\cal E} - (\mathbf{E}^* \times \mathbf{B}^*)_x] 
+ \lambda^* [P^* -  (E^*_x)^2 -(B^*_x)^2].
\label{eq:FhllSx}
\end{equation}
Subtracting Eq.\,(\ref{eq:FhllSx}) from Eq.\,(\ref{eq:Shll}) we arrive
at a quadratic equation for $\lambda^*$
\begin{equation}
a(\lambda^*)^2 + b\lambda^* + c=0 ,
\label{eq:lambda*}
\end{equation}
whose coefficients are
\begin{equation}
\begin{aligned}
a&=F^{hll}_{\cal E} - (\mathbf{E}^* \times \mathbf{B}^*)_x,\\
b&= (\mathbf{E}_{\bot}^*)^2 + (\mathbf{B}_{\bot}^{*})^2 - {\cal E}^{hll}  - F^{hll}_{S^x},\\
c&=S_x^{hll} -  (\mathbf{E}^* \times \mathbf{B}^*)_x,
\end{aligned}
\label{eq:abc}
\end{equation}
where $(\mathbf{B}_{\bot}^*)=(0, B^*_y, B^*_z)^T$ and
$\mathbf{E}_{\bot}^*=(0, E^*_y, E^*_z)^T$. 

In the ideal limit, in which the electric field simply is $\mathbf{E}=-\mathbf{v}\times \mathbf{B}$, the coefficients of
the quadratic equation (\ref{eq:abc}) reduce to the expressions in \cite{Mignone_Bodo:2006}. Thus, employing the same
arguments as in the former paper \citep[see also][]{Mignone:2005MNRAS}, of the two roots of Eq.\,(\ref{eq:lambda*}),
only the one with the minus sign is compatible with $\lambda_l\le \lambda^* \le \lambda_r$ and, therefore, it is the
physically admissible solution.

If the solution resulting from Eq.\,(\ref{eq:lambda*}) is $\lambda^*\ne
0$, then $P^*$ can be recovered from Eq.\,(\ref{eq:FhllSx})
\begin{equation}
\begin{aligned}
P^*   &=  F^{hll}_{S^x} + (E^*_x)^2 + (B^*_x)^2 - \lambda^* [ F^{hll}_{\cal E} - (\mathbf{E}^* \times \mathbf{B}^*)_x].
\end{aligned}
\label{eq:P*}
\end{equation}
In the complementary case that $\lambda^*=0$, using the second RH
condition of (\ref{eq:RH}) together with the flux consistency
condition (\ref{eq:lambda*Fhll}) allows us to express all the fluxes
in terms of the HLL flux (i.e., in terms of the variables in the left
and right states):
\begin{equation}
\mathbf{F}^*_r = \mathbf{F}^*_l = \mathbf{F}^{hll} \quad \text{ (if } \lambda^*=0),
\label{eq:F*_lambda*=0}
\end{equation}
and the calculation of $P^*$ is not necessary to obtain the numerical fluxes.

The energy density ${\cal E}$ (Eq.\,\ref{eq:calE}), can also be expressed
as 
\begin{eqnarray*}
{\cal E} = (S_x - (\mathbf{E} \times \mathbf{B})_x)/v_x - P + \mathbf{E}^2 + \mathbf{B}^2, 
\end{eqnarray*}
if $v_x\ne 0$. In any of the intermediate states, \maa{using Eq.\,(\ref{eq:Ua*_1}), we find:} 
\begin{equation}
\begin{aligned}
{\cal E}_{a}^* =& \frac{1}{\lambda^*}\left[\frac{ \lambda_a S_{x}^{hll} -  \lambda^* F_{\cal E}^{hll}} {\lambda_a-\lambda^*} 
- (\mathbf{E}^* \times \mathbf{B}^*)_x\right] - P^*+\\ & (\mathbf{E}^{*})^2 +(\mathbf{B}^{*})^2 , \\
\end{aligned}
\label{eq:calE*}
\end{equation}
if $\lambda^*\ne0$. The complementary case ($\lambda^*=0$) makes use
of Eq.\,(\ref{eq:F*_lambda*=0}), and the evaluation of
${\cal E}_{a}^*$ is not needed.
\subsection{Discussion of the assumptions made to build the HLLC solver}
\label{sec:HLLCdiscussion}
Notice that in equation Eq.\,\ref{eq:continuity_ContactWave1}, we have
assumed that the speed of the contact wave is equal to the (average)
normal velocity over the Riemann fan. We point out that one may choose
different sets of variables to be continuous across the contact
wave. Our choice is similar (but not equal) to that of
\cite{Mignone_Bodo:2006}. However, we do not impose continuity of the
components of the velocity parallel to the contact wave.  In this
sense, our approach is similar to that of \cite{Li:2005} in classical
MHD or to the one of \cite{Honkkila_Janhunen:2007} or
\cite{Kim_Balsara:2014} in RMHD. Instead, we impose continuity of the
electric field due to its duality with respect to the magnetic field
in RRMHD.

\maa{We have also imposed continuity of the scalar potentials
  $\phi_a^*$ and $\psi_a^*$ and left unspecified the form of their
  corresponding fluxes, $F^*_{\phi,a}$ and $F^*_{\psi,a}$,
  respectively, which are unknowns of the problem.  If we did not
  assume continuity of $\phi^*$ and $\psi^*$, this would be
  incompatible with the RH conditions across the middle wave
  (Eq.\,\ref{eq:RH}) if we also enforced that $F^*_{\phi,a} = B^*_x$
  and that $F^*_{\psi,a} = E^*_x$. Alternatively, we could relax the
  conditions $\phi^*_r = \phi^*_l $ and $\psi^*_r = \psi^*_l$ and
  assume that $F^*_{\phi,a} = B^*_x$ and $F^*_{\psi,a} = E^*_x$, but
  then the resulting system of equations becomes overdetermined. In
  the practical applications where we have used the new HLLC sover,
  this is not a problem though. The reason is that if we assume that
  the form of the fluxes of $\phi_a^*$ and $\psi_a^*$ is
  $F^*_{\phi,a} = B^*_x$ and $F^*_{\psi,a} = E^*_x$, we obtain from
  the RH conditions across the middle wave that $\phi^*_r = \phi^*_l $
  and $\psi^*_r = \psi^*_l$. But then, we have two alternative
  expressions for the fluxes of the latter variables: from the
  consistency relations across the outermost waves
  (Eqs.\,\ref{eq:Uhll} and \ref{eq:lambda*Fhll}), we have
  $F^*_{\phi,a} =F^{\rm hll}_\phi$ and
  $F^*_{\psi,a} =F^{\rm hll}_\psi$, while from the assumed form of the
  fluxes, we obtain $F^*_{\phi,a} =B^{\rm hll}_x$ and
  $F^*_{\psi,a} =E^{\rm hll}_x$, respectively. These alternative
  expressions for the fluxes are in general incompatible. There is,
  however, a possibility to make all these alternative expressions of
  the fluxes in the starred region compatible if and only if
  $\lambda_l=-1$ and $\lambda_r=+1$. Fortunately, this is the default
  choice that we make for the bounds on the limiting speeds of the
  RRMHD system. However, this procedure leaves no choice on the
  selection of $\lambda_l$ and $\lambda_r$, as it is typically the
  case in other HLLC solvers.
  Another way to circumvent the incompatibility found above (if we do
  not enforce $\phi^*_r = \phi^*_l $ and $\psi^*_r = \psi^*_l$) is to
  assume that the fluxes of $\phi^*$ and $\psi^*$ are free
  variables. However, to close the system of equations, we would need
  to impose two additional relations among the fluxes of $\phi^*$ and
  $\psi^*$ on each side of the middle wave. In this regard, two
  considerations are in order. First, the scalar potentials are
  introduced in the algorithm to preserve the constraints
  $\nabla \cdot \mathbf{E}=q$ and $\nabla \cdot \mathbf{B}=0$ and,
  second, we assume that both $\mathbf{E}$ and $\mathbf{B}$ are
  continuous across the middle wave. Taking into account these two
  considerations, we have not found satisfactory alternatives to the
  assumption of continuity of the $\phi^*$ and $\psi^*$ across the
  contact wave in the standard framework of an HLLC solver for {\em
    the whole} set of conserved variables (but see
  App.\,\ref{sec:HLLCa}).}

\section{Numerical Experiments}
\label{sec:tests}

In this section, we demonstrate the performance of the new HLLC solver
in a number of 1D Riemann problems and 2D standard (R)RMHD tests. To
compare our results with the ones in the ideal limit, we set the
conductivity to a large constant value, i.e.\, $\sigma = 10^6$ (see
the discussion in Sec.\,\ref{sec:st2} as well as
Fig.\,\ref{fig:ST2-r}).  All 1D test problems are performed in a
computational domain $x \in \left[ 0, 1\right]$ with and the CFL
factor $\CFL = 0.1$.  The initial discontinuity is placed at $x=
0.5$. RRMHD equations are integrated using the second order MIRK
method \citep{Aloy_Cordero-Carrion:2016}.  However, we note that most
of the results presented in this paper are computed employing a
(globally) $1^{\rm st}$-order accurate scheme, resulting from not
applying any intercell reconstruction to the numerical variables
(Godunov method). In this way, a comparison of the performance of the
HLL and HLLC Riemann solvers can be shown more clearly.  The initial
conditions for 1D tests are summarised in Tab.\,\ref{tab:models}.
\begin{table*}
\centering 
\caption{Initial conditions for 1D test problems.
  In all simulations, we set the pseudopotentials and the charge
  density to zero, i.e.\,$\psi = \phi = q = 0$,  and the electric
  field to its ideal approximation $\mathbf{E} = - \mathbf{v} \times
  \mathbf{B}$.
  The columns give the test name, the state (left; L or right; R),
  rest-mass density $\rho$, gas pressure $p_{\rm g}$, velocity and magnetic field
  components ($\mathbf{v}$ and $\mathbf{B}$, respectively),  the adiabatic index $\gamma$, the simulation time $t$, and
  the number of zones.}
\begin{tabular}{ ccccccccccccc  }
Test   &  State & $\rho$  & $p_g$  & $v_x$   & $v_y$  & $v_z$  & $B_x$ & $B_y$   & $B_z$  & $\gamma$ & t  & Zones   \\ \hline
CW1    &  L     & $10.0$  & $1.0$  & $0.0$   & $0.7$  &$0.2$  & $5.0$ & $1.0$   & $0.5$  & $5/3$   & 1.0   & 40 \\
       &  R     & $1.0$   & $1.0$  & $0.0$   & $0.7$  &$0.2$  & $5.0$  & $1.0$   & $0.5$  &         &&       \\ \hline
CW2    &  L     & $1.0$   & $1.0$  & $0.2$   & $0.0$  &$0.0$  & $1.0$ & $1.0$   & $0.0$  & $5/3$   & 1.0  & 40 \\
       &  R     & $0.125$ & $1.0$  & $0.2$   & $0.0$  &$0.0$  & $1.0$  & $1.0$   & $0.0$  &         &   &    \\ \hline
RW    &  L     & $1.0$   & $1.0$  & $0.4$   & $-0.3$  &$0.5$  & $2.4$ & $1.0$   & $-1.6$  & $5/3$   & 1.0  & 40 \\
       &  R     & $1.0$ & $1.0$  & $0.377237$   & $-0.482389$  &$0.424190$  & $2.4$  & $-0.1$   & $-2.178213$  &         &   &    \\ \hline
ST1    &  L     & $1.0$   & $1.0$  & $0.0$   & $0.0$  &$0.0$  & $0.5$  & $1.0$   & $0.0$  & $2.0$   & 0.4  &400 \\
       &  R     & $0.125$ & $0.1$  & $0.0$   & $0.0$  &$0.0$  & $0.5$  & $-1.0$  & $0.0$  &         &      & \\ \hline
ST1-B0 &  L     & $1.0$   & $1.0$  & $0.0$   & $0.0$  &$0.0$  & $0.0$  & $1.0$   & $0.0$  & $2.0$   & 0.4  &400 \\
       &  R     & $0.125$ & $0.1$  & $0.0$   & $0.0$  &$0.0$  & $0.0$  & $-1.0$  & $0.0$  &         &       &\\ \hline
ST2    &  L     & $1.08$  & $0.95$ & $0.4$   & $0.3$  &$0.2$  & $2.0$  & $0.3$   & $0.3$  & $5/3$   & 0.55 &800 \\
       &  R     & $1.0$   & $1.0$  & $-0.45$ & $-0.2$ &$0.2$  & $2.0$  & $-0.7$  & $0.5$  &         &       &\\ \hline
ST3    &  L     & $1.0$   & $0.1$  & $0.999$ & $0.0$  &$0.0$  & $10.0$ & $7.0$   & $7.0$  & $5/3$   & 0.4   &400\\
       &  R     & $1.0$   & $0.1$  & $-0.999$& $0.0$  &$0.0$  & $10.0$ & $-7.0$  & $-7.0$ &         &       &\\ \hline
ST4    &  L     & $1.0$   & $5.0$  & $0.0$   & $0.3$  &$0.4$  & $1.0$  & $6.0$   & $2.0$  & $5/3$   & 0.5   &800\\
       &  R     & $0.9$   & $5.3$  & $0.0$   & $0.0$  &$0.0$  & $1.0$  & $5.0$   & $2.0$  &         &       \\ \hline
ST5    &  L     & $1.0$   & $30.0$ & $0.0$   & $0.0$  &$0.0$  & $5.0$  & $6.0$   & $6.0$  & $5/3$   & 0.4   &800\\
       &  R     & $1.0$   & $1.0$  & $0.0$   & $0.0$  &$0.0$  & $5.0$  & $0.7$   & $0.7$  &         &       &\\ \hline
\end{tabular}
\label{tab:models}
\end{table*}
\begin{table*}
\centering 
\caption{Mean ratios of CPU running time between the HLLC solver and the
  HLL solver, $t_{\rm hllc}/t_{\rm  hll}$. To compute the mean ratios in 1D
  tests (ST1 to ST5), we first measure the ratio of CPU running times
  for each of the working resolutions employed in this paper, $t_{\rm
    hllc}/t_{\rm  hll}(n_{x,i})$, $n_{x,i} = \{50, 100, 200, 400, 800,
  1600, 3200\}$, and then average over all of these results (i.e.,
  $t_{\rm hllc}/t_{\rm  hll} xs:=\frac{1}{7}\sum_{i=1}^7 t_{\rm
    hllc}/t_{\rm  hll}(n_{x,i})$). Likewise, for the 2D tests (RR and CE)
  the average of ratios of CPU time is computed performing them with
  resolutions of $50 \times 50$, $100 \times 100$,
  $200 \times 200$ and $400 \times 400$.}
\begin{tabular}{ ccccccccc }
\multicolumn{9}{c}{\textbf{CPU running times}}                                         \\  \hline \hline
test              &  ST1   & ST1-B0   & ST2    & ST3   & ST4  & ST5   & RR    & CE     \\ \hline
$t_{\rm hllc}/t_{\rm hll}$  &  $1.06$ & $1.06$  & $1.06$ & $1.13$ & $1.18$ & $1.15$ & $1.05$ &  $1.18$ \\ \hline 
\end{tabular}
\label{tab:ct}
\end{table*}

\subsection{Contact Wave Discontinuities (CW1, CW2)}
\label{sec:CW}

\begin{figure*}
  \centering
  \scalebox{0.60}{\hspace{-2cm}\input{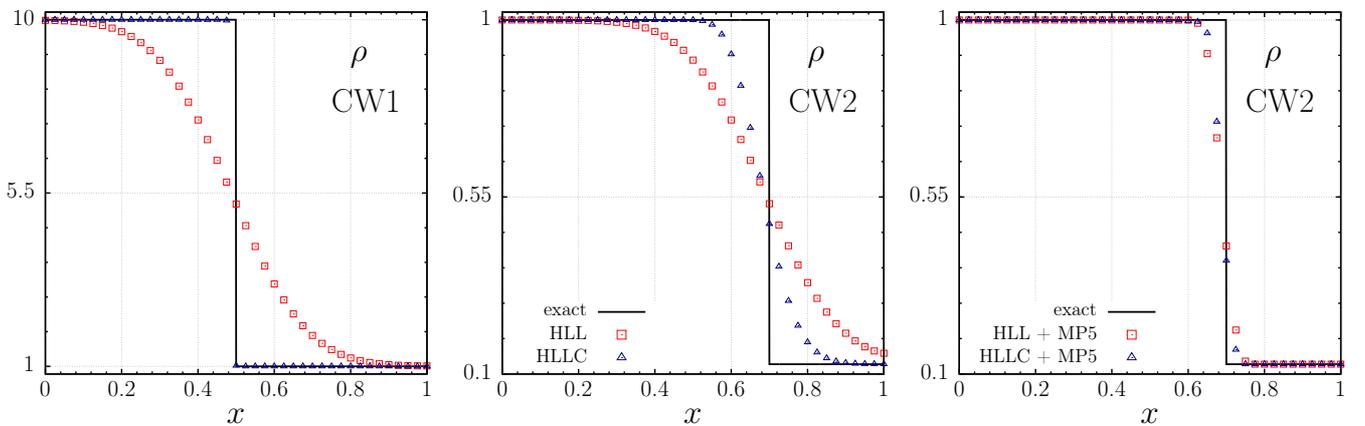}}
  \caption{ \textbf{CW:} Rest-mass density at $t = 1.0$ in contact
    wave discontinuity tests CW1 (left panel) and CW2 (central and
    right panels; see Tab.\,\ref{tab:models}) performed with the HLL
    (red squares) and the HLLC (blue triangles) approximate Riemann
    solvers.  The analytical solution is depicted with a black solid
    line.  In the simulation presented in the right panel, the MP5
    scheme is used, whereas in the other simulations the piecewise
    constant reconstruction (Godunov) is employed.}
  \label{fig:CW}
\end{figure*}

Contact wave discontinuity tests (CW1 and CW2 in
Tab.~\ref{tab:models}) consist in simulating an isolated contact
discontinuity with a jump only in the mass density $\rho$.  In the CW1
test, where $v_x=0$, the HLLC solver does not show any smearing of the
initial profile, as for this particular case, HLLC resolves the
discontinuity (Fig.\,\ref{fig:CW} left panel) exactly. This behaviour
is expected for the HLLC solver, which is specifically built to
include a contact wave in the numerical solution. In the CW2 test,
whose setup was proposed by \cite{Honkkila_Janhunen:2007}, there is a
smearing of the contact wave, which is smaller for the HLLC solver
than for the HLLC solver (Fig.\,\ref{fig:CW} central panel). The
smearing of the contact wave is a result of two facts: ($i$) the
non-Lorentzian invariance of the numerical viscosity added by the HLL
and HLLC solvers; ($ii$) an inherent numerical diffusion of
shock-capturing methods of Godunov-type \cite[unless specific
techniques to track internal interfaces dynamics are employed; see,
e.g.,][]{Abgrall_Karni:2001}.

We have considered several variants of the CW2 test where
progressively larger values of $v_x$ are taken. As $v_x\rightarrow 1$,
the diffusion of the contact discontinuity becomes larger in both
approximate Riemann solvers and the differences between them reduce
(yet the HLLC solver always displays smaller smearing than the HLL
solver). This is due to the degeneration of the Riemann structure when
any of the components of the $3$-velocity is close to the speed of
light. Under such conditions, all eigenvalues of the system of RRMHD
equations approach the speed of light and the contact wave is resolved
as a shock by the algorithm (thus, unphysical numerical dissipation is
added).

\subsection{Rotational Wave (RW)}
\begin{figure}
  \centering
  \scalebox{0.55}{\input{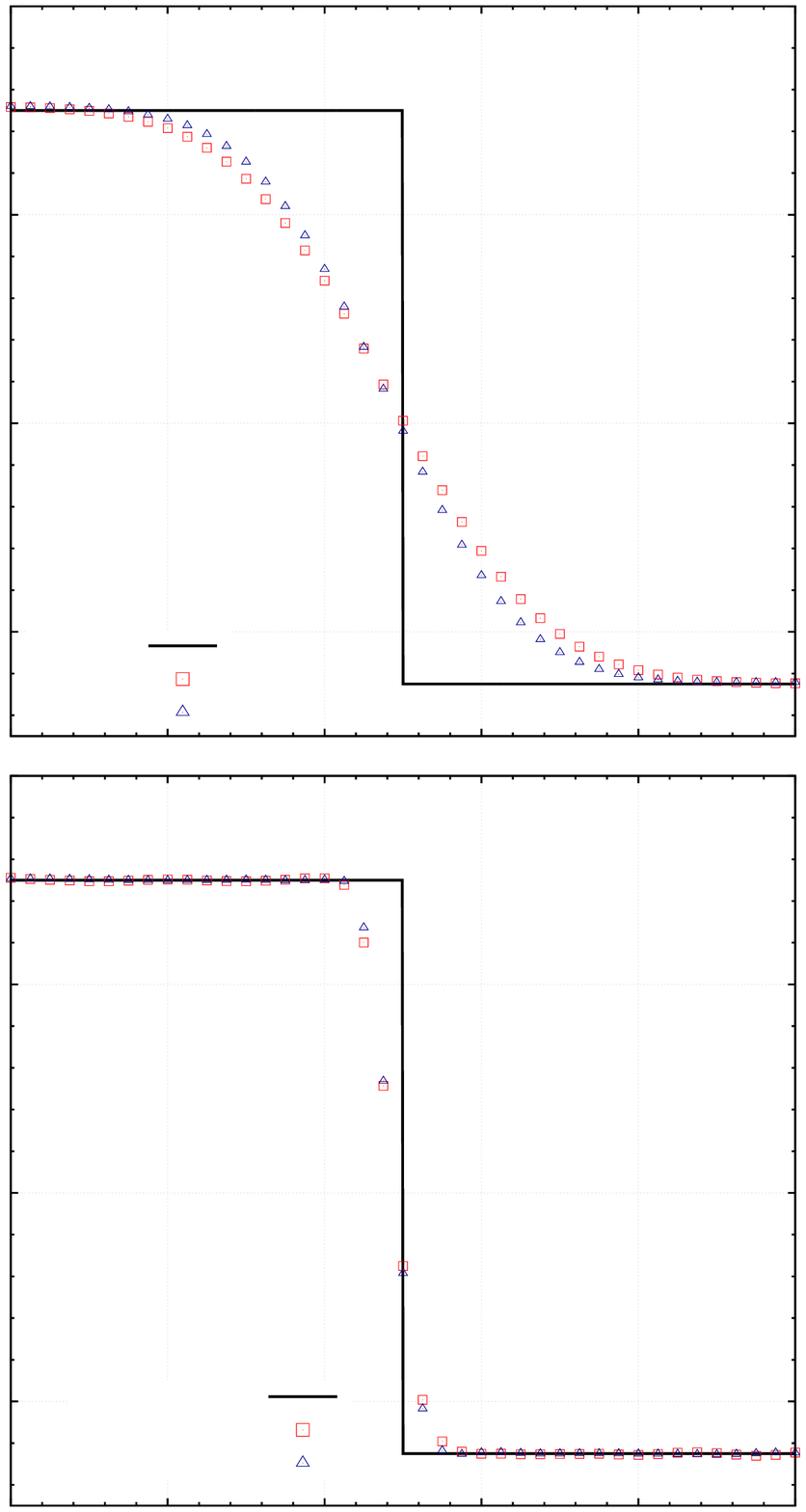}}
  \caption{\textbf{RW:} Magnetic field ($B_y$ component) for an
    isolate rotational wave at $t=1$. The solid line stands for the
    analytic solution, blue triangles and red squares symbols show the
    solution with the HLLC and HLL solvers, respectively. The upper
    panel is computed without employing any high-order intercell
    reconstruction (Godunov), while the lower panel corresponds to the
    same test using the MP5 reconstruction.}
  \label{fig:RW}
\end{figure}

Following \citep{Mignone_etal:2009}, we set up an isolated rotational
wave like in their ideal RMHD test.  Across a rotational wave, the
components of vector fields exhibit jumps (yet their moduli are
preserved), while scalar quantities (e.g., rest-mass density, total
pressure, etc.) are continuous. We show in Fig.\,\ref{fig:RW} the
variation of $B_y$ across the rotational wave. Neither the HLL nor the
HLLC solver is able to capture the initially specified jump (see
Tab.\,\ref{tab:models}) and both smear the solution (the HLL solver
slightly more than the HLLC solver).

As noted in \cite{Anton_etal:2010}, in practical applications, the
numerical diffusion is not as large  as one may infer from
the results of Fig.\,\ref{fig:RW} (upper panel). To ameliorate
the problem, a high-order spatial interpolation (instead of using
simply the Godunov method) is advisable. Thus, we have repeated both
the RW and also the CW2 tests employing a $5^{\rm th}$-order 
monotonicity-preserving (MP5) intercell reconstruction
\citep{Suresh_Huynh_1997}. The results are shown in
Figs.\,\ref{fig:CW} (right panel) and \ref{fig:RW} (bottom panel). In
both cases, only 4 numerical zones are needed to resolve the isolated
discontinuities.

\subsection{Shock Tube Problem 1 (ST1)}
\begin{figure*}
  \centering
  \scalebox{0.725}{\input{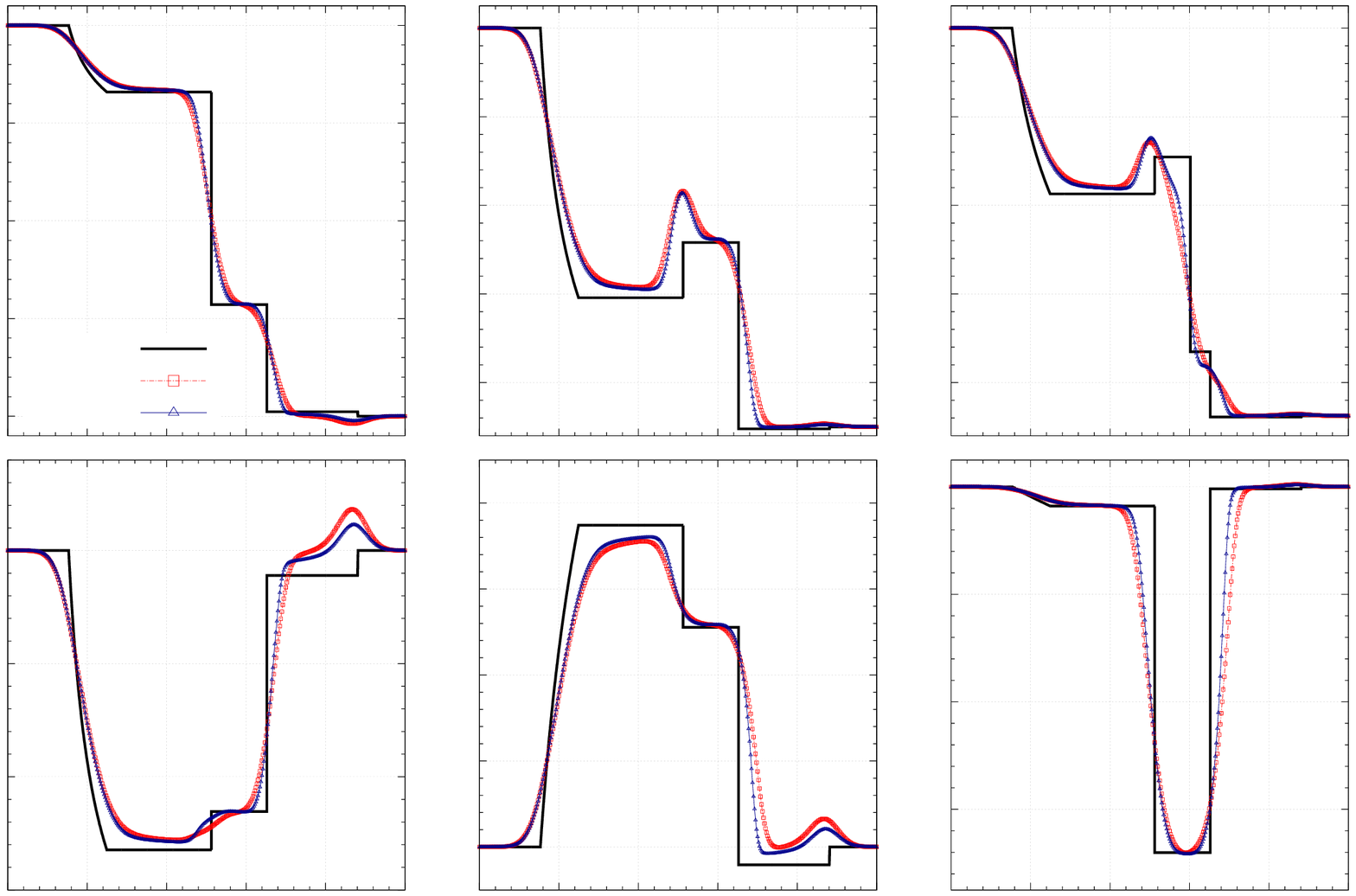}}
  \caption{\textbf{ST1:} Relativistic Brio-Wu shock tube test at
    t=0.4. The calculations are performed with a $1^{\rm st}$-order accurate
    scheme on a grid of 400 zones employing the HLL (red symbols) and
    HLLC (blue symbols) Riemann solvers. The analytic solution of this
    test in ideal RMHD is displayed with black solid line. Upper
    panel: magnetic field ($B_y$ component), thermal pressure ($p$),
    rest-mass density ($\rho$). Lower panel: electric field ($E_z$
    component), velocity field ($v_x$ and $v_y$ components).}
  \label{fig:ST1}
\end{figure*}

\begin{figure*}
  \centering
  \scalebox{0.725}{\input{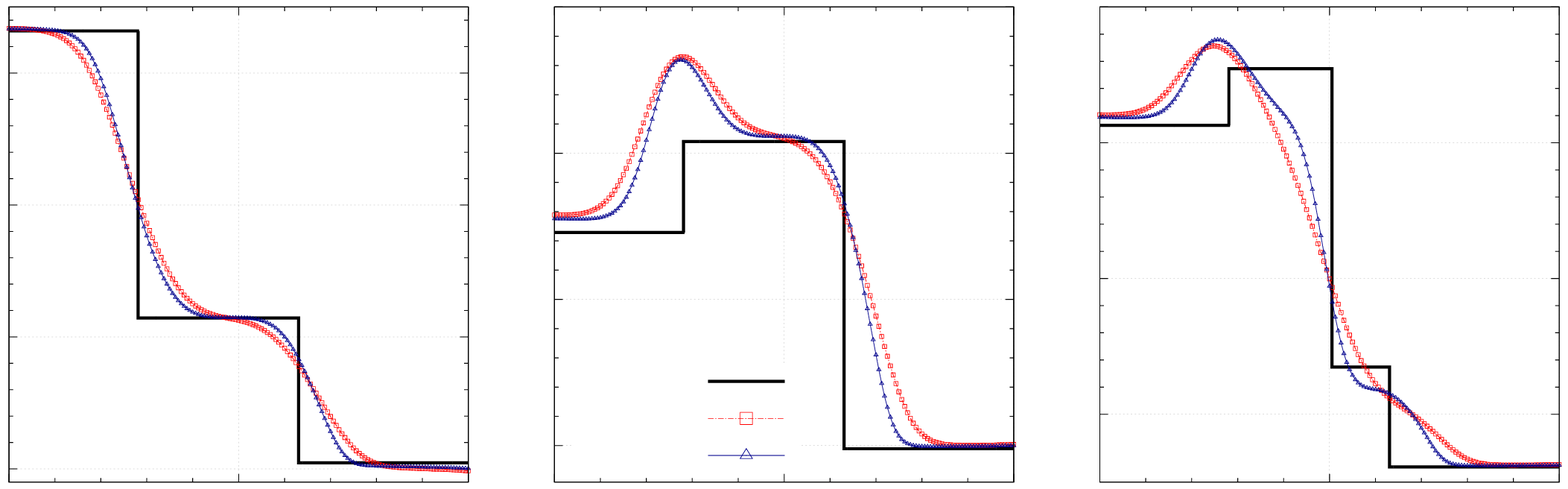}}
  \caption{\textbf{ST1:} Zoom of Fig.\,\ref{fig:ST1} on the region
    where the contact discontinuity is located. From left to right:
    magnetic field ($B_y$ component), thermal pressure ($p_{\rm g}$)
    and rest-mass density ($\rho$).}
  \label{fig:ST1_zoom}
\end{figure*}

The test of \cite{Balsara:2001} \citep[which is a relativistic
extension of the test proposed by][]{Brio_Wu:1988} has been used by a
number of practitioners in the RMHD field
\citep[e.g.,][]{DelZanna_etal:2003,Leismann_etal:2005,Mignone_Bodo:2006,Mignone_etal:2009,Anton_etal:2010}. It
considers a fluid with adiabatic index $\gamma = 2$.  The initial
discontinuity breaks into a left-going fast rarefaction, a left-going
compound wave, a contact discontinuity, a right-going slow shock and a
right-going fast rarefaction wave (Fig.\,\ref{fig:ST1}). According to
the analytic solution
\citep{Giacomazzo_Rezzolla:2006}\footnote{\cite{Giacomazzo_Rezzolla:2006}
  solution is built explicitly neglecting compound waves. Besides
  their physical or unphysical nature, numerically the compound wave
  shows up at $x\simeq 0.5$, inducing the seemingly large deviation of
  the numerical solution from the analytic one in the vicinity of its
  location.}, the contact discontinuity must be located at
$x \approx 0.6$. The structure of the contact is more accurately
resolved by the HLLC solver than by the HLL solver as can be seen in
the zoomed Fig.\,\ref{fig:ST1_zoom}.

Both solvers are slightly more diffusive in this test than their ideal
RMHD counterparts (compare our Fig.\,\ref{fig:ST1_zoom} to, e.g.,
Fig.\,4 of \cite{Mignone_etal:2009} or to Fig.\,4 of
\cite{Anton_etal:2010}, where the test is shown with the same
resolution). This extra diffusivity comes from two sources. First, our
estimates for the limiting speeds of the approximate Riemann solvers
($\lambda_r=+1$, $\lambda_l=-1$; Sec.\,\ref{sec:HLLC1}) are larger in
absolute value than the fast magnetosonic waves for this test in ideal
RMHD. The latter are commonly employed as estimators of the fastest
signal speeds in HLL and HLLC approximate Riemann solvers for
RMHD. Note, however, that fast-magnetosonic waves propagate at the
speed of light in RRMHD. Second, the notorious degeneracy of the
eigenfields limiting the Riemann fan (Eq.\,\ref{eq:lambdaEB})
increases the amount of numerical dissipation added to characteristic
variables corresponding to eigenvalues without jumps for certain
electromagnetic configurations.

\begin{figure*}
  \centering
  \scalebox{0.725}{\input{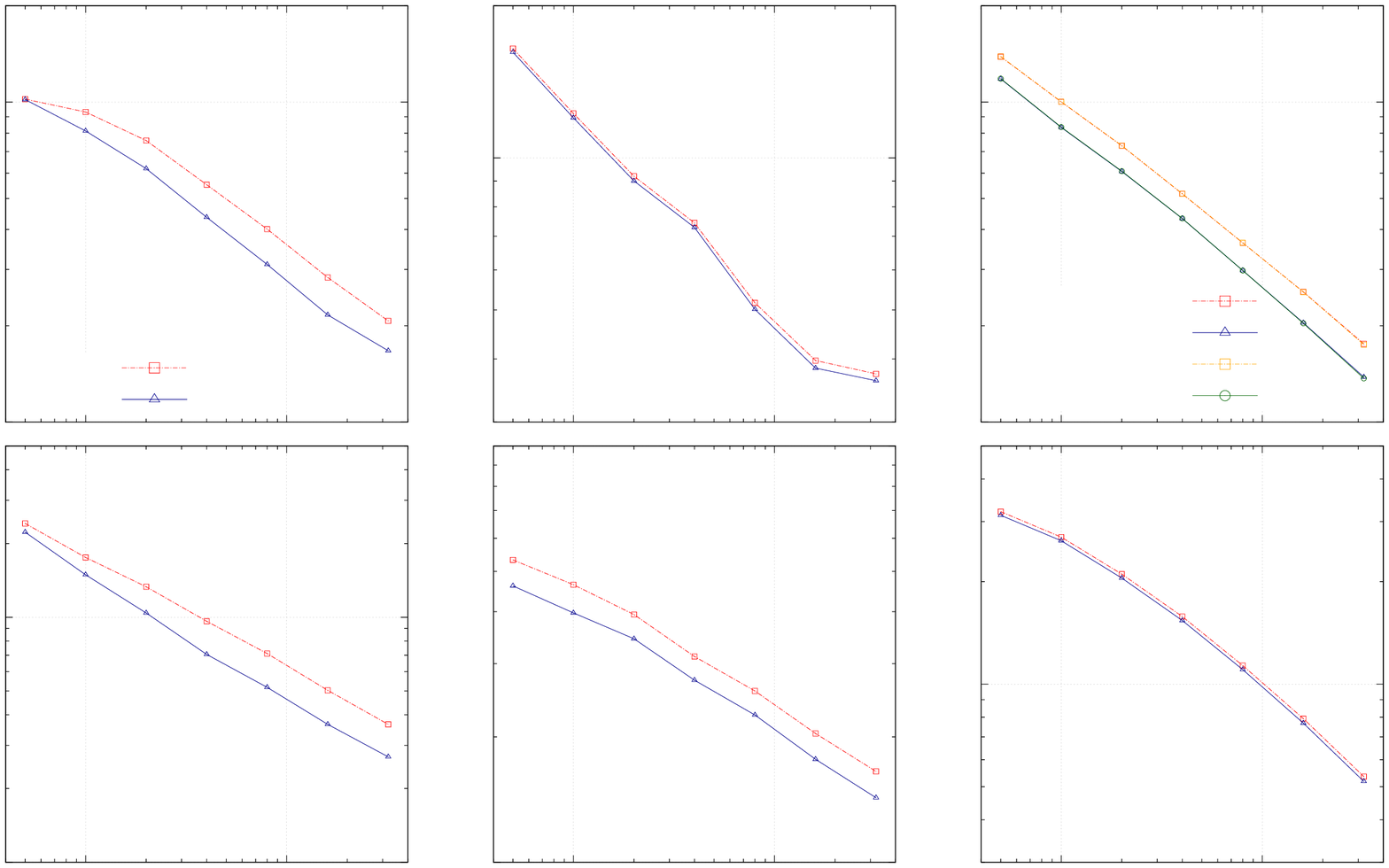}}
  \caption{L1-norm errors (Eq.\,\ref{eq:err})  
of $B_y$ magnetic field component in 
  six shock tube problems presented in 
    Tab.\,\ref{tab:models}. All tests were performed with resolutions of 
$n_x = 50, 100, 200, 400, 800, 1600,  3200$ grid zones, and a 
    small $\CFL = 0.1$ to avoid contributions to the error arising
    from the time discretisation.  The numerical solutions computed with the HLL and HLLC solvers
are displayed with red and blue squares,
    respectively. }
  \label{fig:L1-norm}
\end{figure*}

To quantify the accuracy of the HLL and HLLC solvers, we use the
discrete L1-norm error, in the form:
\begin{equation}
\varepsilon_{\rm L1} = \frac{1}{n_x}\sum_{i=1}^{n_x}  \left| U_i^{\rm ref} - U_i \right|, 
\label{eq:err}
\end{equation}
where $U_i^{\rm ref}$ is a reference solution computed with the exact
solver of \cite{Giacomazzo_Rezzolla:2006} on the same grid as the
numerical solution $U_i$.  $n_x$ is the number of numerical grid zones
in the $x-$direction. For the 1D tests, we evaluate the L1-norm errors
of the $B_y$ component of the  magnetic field
and display them as a function of the number of numerical zones in
Fig.\,\ref{fig:L1-norm}. 

The measured errors for the ST1 test are $\sim 20\%$ smaller using the
HLLC solver than using the HLL solver for $n_x>100$
(Fig.\,\ref{fig:L1-norm}, top left panel), reflecting that the HLLC
solutions are closer to the exact ones than the HLL numerical
approximations.

%
%

\begin{figure*}
  \centering
  \scalebox{0.725}{\input{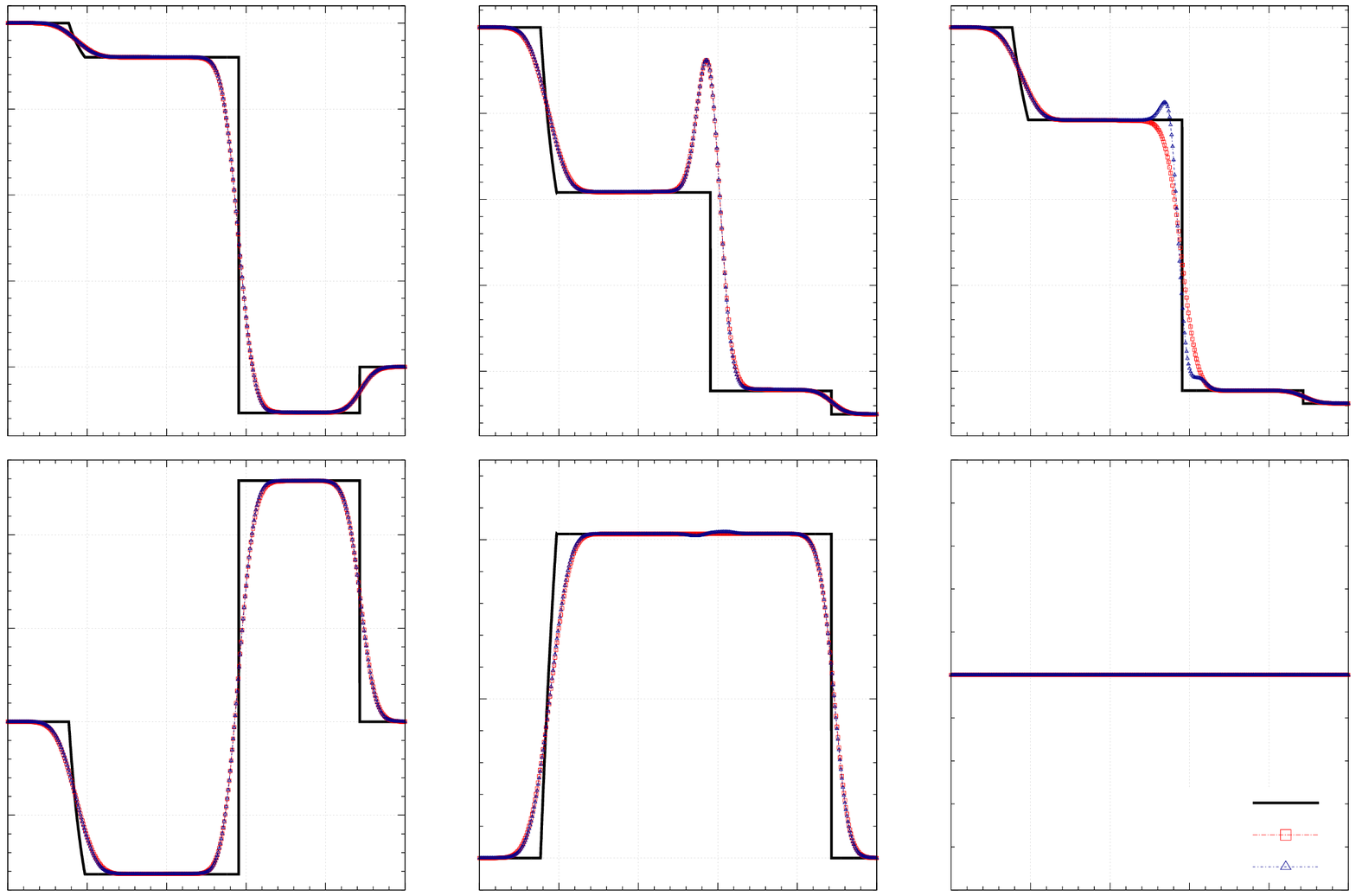}}
  \caption{\textbf{ST1-B0:} Variant of the Relativistic Brio-Wu shock
    tube test at t=0.4 when the longitudinal magnetic field is zero
    ($B_x=0$). The panels, lines and symbols are the same as in Fig.\,\ref{fig:ST1}.}
  \label{fig:ST1-B0}
\end{figure*}

\begin{figure*}
  \centering
  \scalebox{0.725}{\input{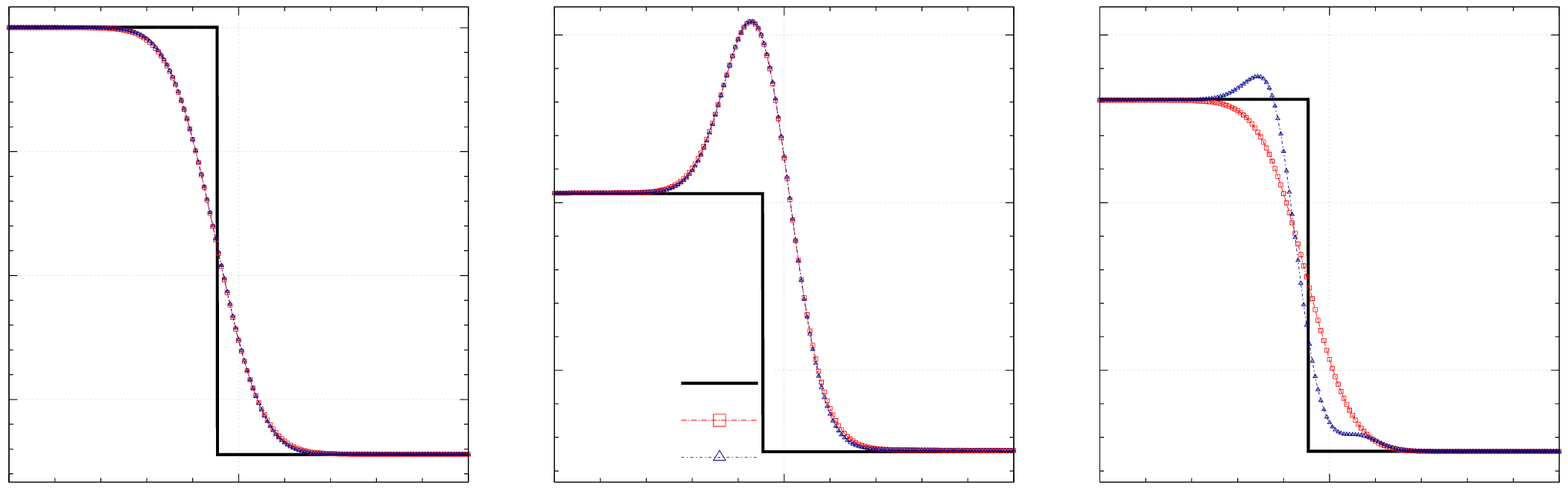}}
  \caption{\textbf{ST1-B0:} Same as Fig.\,\ref{fig:ST1_zoom} but for
    test ST1-B0.}
  \label{fig:ST1-B0_zoom}
\end{figure*}

Differently from the ideal RMHD implementation of the HLLC solver by
\cite{Mignone_Bodo:2006}, our approximate Riemann solver does not need
to distinguish between the cases $B_x=0$ and $B_x\ne 0$ (see
Sec.\,\ref{sec:HLLC1}). To demonstrate this, we consider the ST1-B0
test (see Tab.\,\ref{tab:models}) which is like the ST1 test, but with
$B_x=0$. The structure of the solution to this test only contains two
fast waves, a rarefaction moving to the left and a shock moving to the
right with a tangential discontinuity between them. In contrast to the
standard ST1 test, compound waves are not present in ST1-B0. The new
HLLC solver can deal with this setup at the same quantitative level as
in the case with $B_x\ne 0$ (Fig.\,\ref{fig:ST1-B0}).  A small
overshooting in the thermal pressure across the tangential
discontinuity (at $x\simeq 0.63$) is present in both approximate
Riemann solvers, but the contact wave is more sharply resolved with
the HLLC than with the HLL solver (Fig.\,\ref{fig:ST1-B0_zoom}).
%

Discrete L1-norm errors of the magnetic field $B_y$ component,
computed for the ST1-B0 test, are plotted in the upper mid panel of
Fig.\,\ref{fig:L1-norm}. In this case the difference between L1-norm
errors for the HLL and HLLC solvers reduces substantially as compared
to the ST1 test. This happens because the jump accros the contact wave
is zero in this test and, as a result, the potential advantage of the
HLLC solver with respect to the HLL solver drastically
reduces. Anyway, the HLLC L1-norm errors are a bit smaller than the
corresponding HLL L1-norm errors at all resolutions. We observe that
the computational overhead of the HLLC solver with respect to the HLL
solver is very modest (Tab.\,\ref{tab:ct}). On average, employing the
former solver takes only $\sim 6\%$ more time to compute the solution
in the ST1 and ST1-B0 tests.


\subsection{Shock Tube Problem 2 (ST2)}
\label{sec:st2}
\begin{figure*}
  \centering
  \scalebox{0.725}{\input{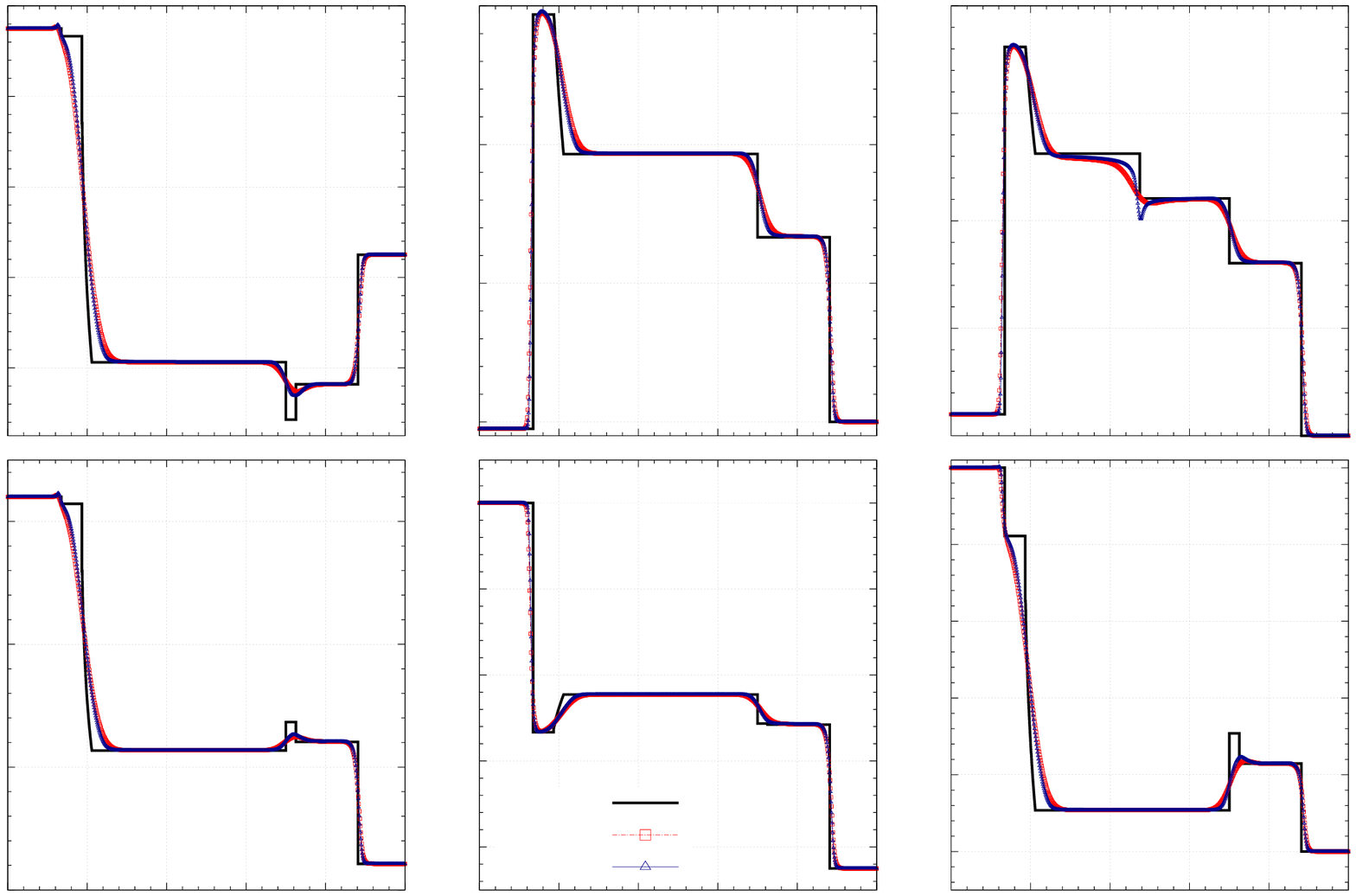}}
  \caption{\textbf{ST2:} Second 1D Riemann problem
    (Tab.\,\ref{tab:models}) after $t=0.55$ and using 400 numerical
    zones.  The variables displayed and the meaning of the symbols and
    lines is the same than in Fig.\,\ref{fig:ST1}.}
  \label{fig:ST2}
\end{figure*}

\begin{figure*}
  \centering
  \scalebox{0.725}{\input{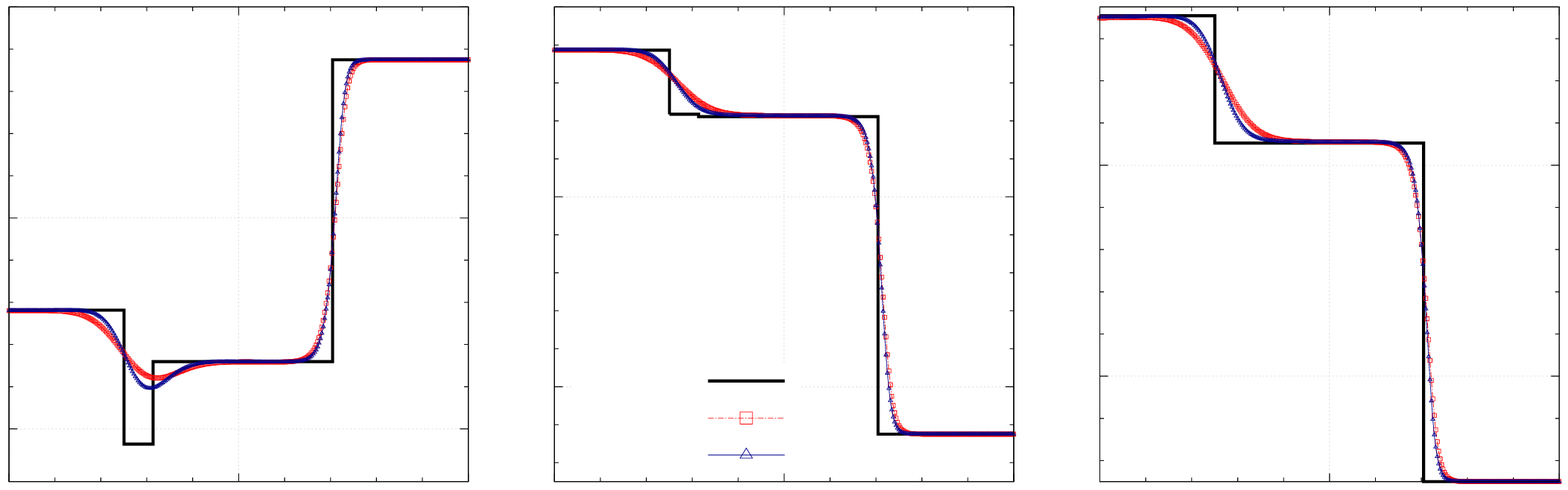}}
  \caption{\textbf{ST2:} Zoom of some of the variables displayed in
    Fig.\,\ref{fig:ST2} focusing on the right-going slow shock
    ($x \simeq 0.7$), Alfv\'en wave ($x\simeq 0.73$) and fast shock
    ($x\simeq 0.88$). From left to right: magnetic field ($B_y$
    component), velocity ($v_x$ component) and rest-mass density
    ($\rho$).}
  \label{fig:ST2_zoom}
\end{figure*}

In this shock tube problem, the break up of the initial discontinuity
develops a Riemann fan with 7 different waves in the ideal limit. We
use the same set up as \cite{Anton_etal:2010} (in RMHD) or as
\cite{Dumbser_Zanotti:2009} or \citet{Bucciantini_DelZanna:2013} in
RRMHD, where the computational domain, $x \in [0,1]$, is resolved with
a uniform grid of $400$ cells and the test is evolved until a final
time $t = 0.55$. The HLLC solver attains a sharper representation of
the entropy wave located at $x\simeq 0.48$ than the HLL solver
(Fig.\ref{fig:ST2}). Nonetheless, an undershooting ahead of the
entropy wave is evident in the HLLC solution. The state between the
right-going slow shock ($x\simeq 0.7$) and the right-going Alfv\'en
wave at $x\simeq 0.73$ is smeared by both solvers albeit the HLLC
solution is slightly closer to the analytical one also in that region
(Fig.\,\ref{fig:ST2_zoom}). It is also the case for the state between
the left-going Alfv\'en wave ($x\simeq 0.185$) and the left going slow
rarefaction at $x\simeq 0.19$. In the latter case, the two waves are
so close to each other that none of the approximate Riemann solvers
can resolve them. Indeed, this is to be expected, since resolving this
region is a challenge even for the exact solver of
\cite{Giacomazzo_Rezzolla:2006}, which cannot obtain a solution for
this test with an accuracy better than $3.4\times 10^{-4}$. 

The L1-norm errors in this test (Fig.\,\ref{fig:L1-norm} upper right
panel) show again a larger accuracy of the solutions computed with
the HLLC solver. These errors are $\sim 10\%$ lower than for
equivalent tests performed with the HLL solver, almost independently
of the resolution employed. Remarkably, for the ST2 test, the
computational cost to obtain the solution employing the HLLC solver is
only a $\sim 6\%$ larger than using the HLL solver
(Tab.\,\ref{tab:ct}).

\begin{figure}
  \centering
  \scalebox{0.65}{\input{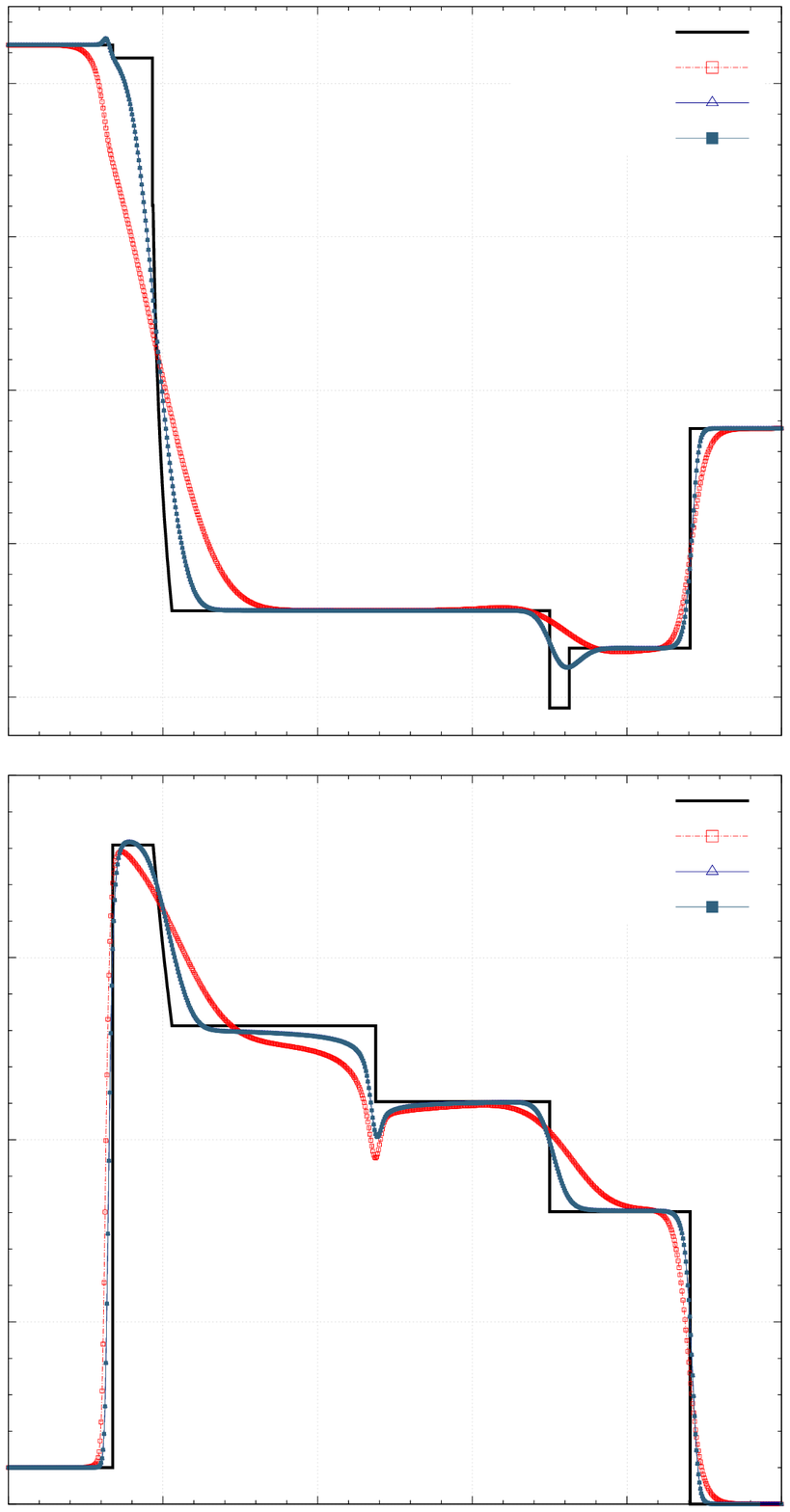}}
  \caption{\textbf{ST2-resistive:} Results of the second 1D Riemann
    problem (ST2; Tab.\,\ref{tab:models}) after $t=0.55$ and using 800
    numerical zones, but for different conductivities
    $\sigma=10^9, 10^6, 10^2$ and the HLLC solver.  Note that the
    lines and symbols corresponding to the case $\sigma=10^9$ overlap
    with the case $\sigma=10^6$. }
  \label{fig:ST2-r}
\end{figure}

In Fig.\,\ref{fig:ST2-r}, we explore the dependence of the results on
the conductivity $\sigma$ analogously to the studies done by
\cite{Dumbser_Zanotti:2009} and
\cite{Bucciantini_DelZanna:2013}. Resistive versions of the shock tube
problem ST2 display smoother profiles of $B_y$ with decreasing
conductivity. Noteworthy, the cases with $\sigma=10^9$ and $10^6$
basically overlap at a resolution of $n_x=800$ zones.  From this fact,
we draw two conclusions.  First, since the results basically are
insensitive to the exact value of the conductivity, when it is large
enough, we can assess that for $\sigma\gtrsim 10^6$ the numerical
resistivity is larger than the physical one in this particular set
up. Second, the {\em proximity} of the results to the analytic
solution justifies our choice of $\sigma=10^6$ as a conductivity value
close enough to the ideal RMHD limit.
This result is expressed more quantitatively in
Fig.\,\ref{fig:L1-norm} (upper right panel; orange and green
lines). For both solvers, the L1-norm errors are basically the same in
tests set up with $\sigma=10^6$ and $\sigma=10^9$.
\subsection{Shock Tube Problem 3 (ST3)}
\begin{figure*}
  \centering
  \scalebox{0.725}{\input{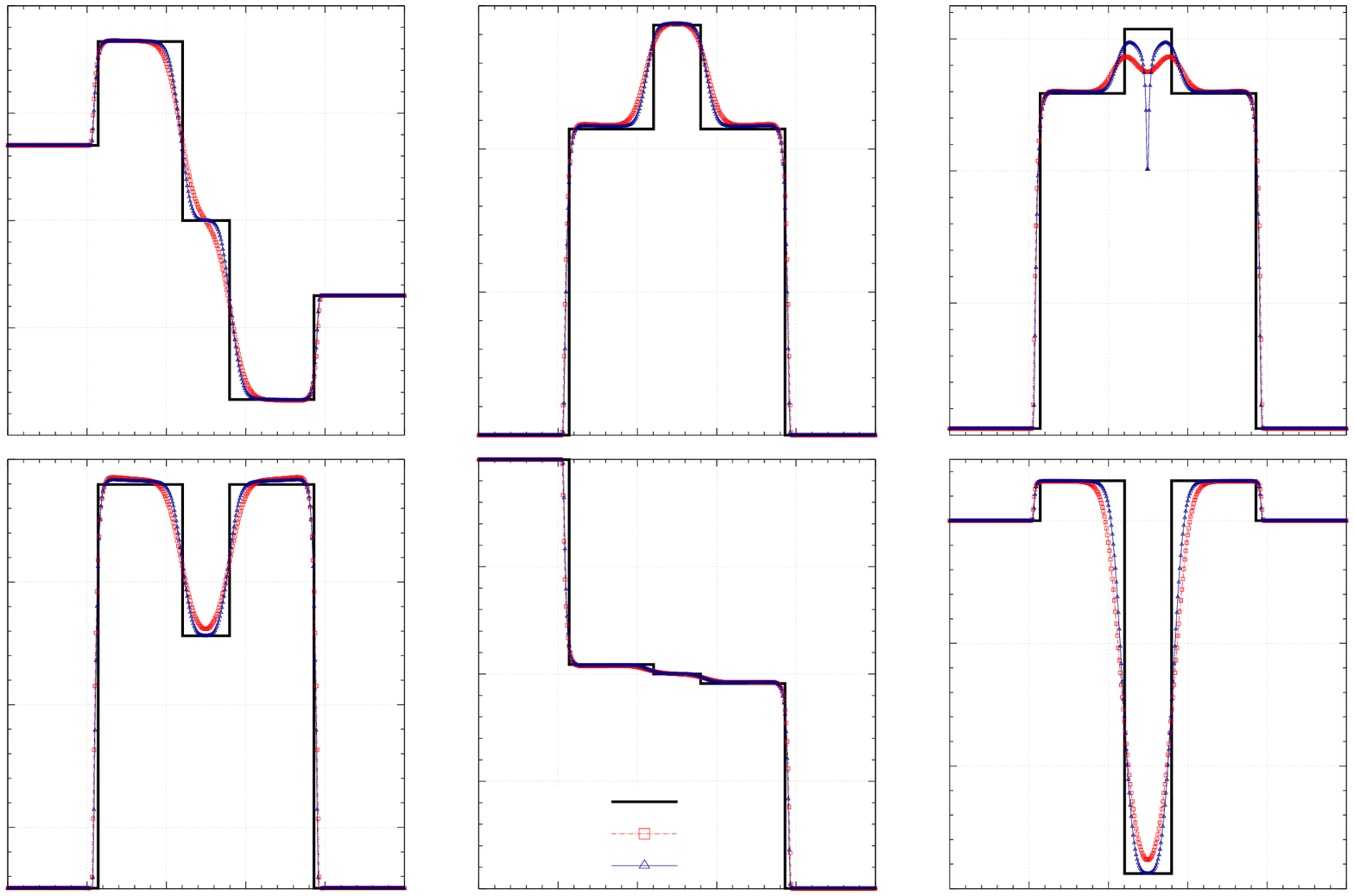}}
  \caption{\textbf{ST3:} Collision of two oppositely moving streams of
    plasma after $t=0.4$ computed with 400 numerical zones. Upper
    panels (from left to right): magnetic field ($B_y$ component),
    thermal pressure ($p_{\rm g}$), rest-mass density ($\rho$). Lower
    panels (from left to right): electric field ($E_z$ component),
    velocity ($v_x$ and $v_y$ components).}
  \label{fig:ST3}
\end{figure*}

In this test, two relativistic, magnetised streams collide producing
two ``reverse'' strong relativistic fast shocks propagating
symmetrically from the mid point of the computational domain
\citep[see,
e.g.,][]{Balsara:2001,DelZanna_etal:2003,Leismann_etal:2005,Mignone_Bodo:2006,Mignone_etal:2009,Anton_etal:2010}.
Following the fast shocks, a pair of symmetric slow shocks further
thermalise the plasma in the state delimited by them, converting all
the remaining kinetic energy into thermal energy and leaving the fluid
at rest (Fig.\,\ref{fig:ST3}). In this test there is not a jump of the
variables across the contact wave (standing at $x=0$) and, therefore,
the differences between HLL and HLLC are rather small. However, the
HLLC approximate Riemann solver produces a slightly sharper
representation of the intermediate discontinuities (slow shocks).
\begin{figure*}
  \centering
  \scalebox{0.725}{\input{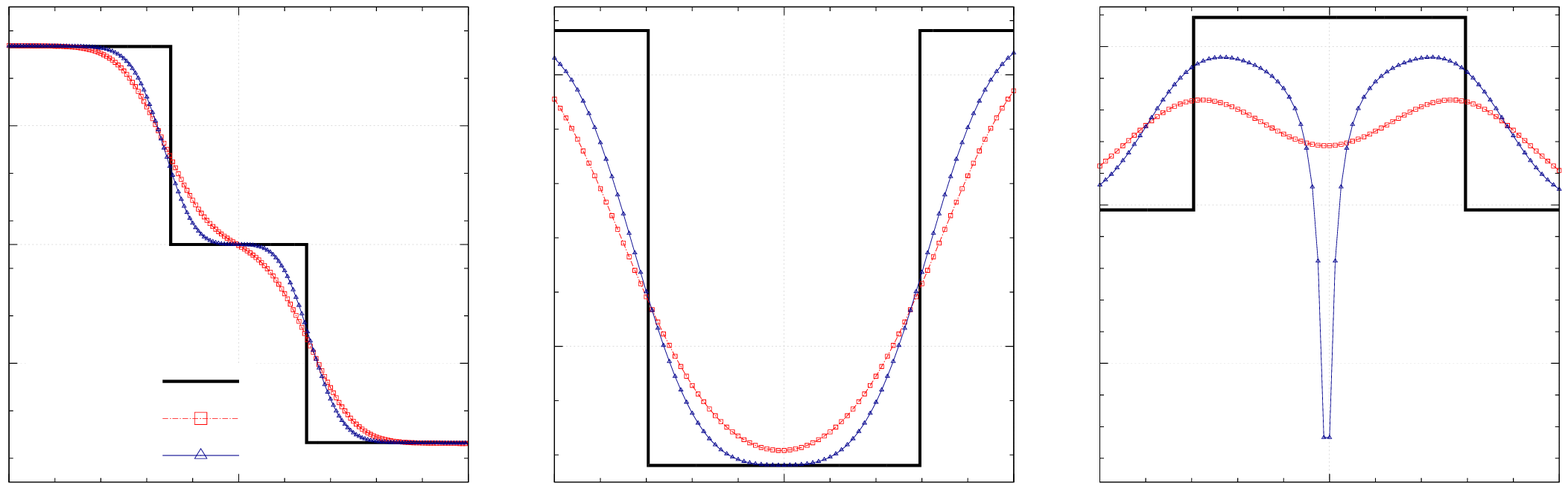}}
  \caption{\textbf{ST3:} Magnification of the central region of the
    ST3 test encompassing the two slow shocks located at
    $x\simeq 0.44$ and $x\simeq 0.56$. From left to right: magnetic
    field ($B_y$ component), velocity ($v_y$ component) and rest-mass
    density ($\rho$).}
  \label{fig:ST3_zoom}
\end{figure*}

Even though both solvers display a pathological undershooting of the
density at the grid center, the density decrease for the HLLC solver
is substantially larger than for the HLL solver. This is a well know
pathology of many Godunov-type schemes know as ``wall heating''
problem \citep{Noh:1987}, which is more stringent for approximate
solvers possessing smaller numerical dissipation. Hence, the HLL
solver displays a much smaller undershooting than the HLLC solver at
$x=0.5$ (Fig.\,\ref{fig:ST3_zoom}). The slightly higher quality of the
HLLC solution is quantified by the smaller L1-norm errors obtained
with the latter solver in comparison to the HLL solver
(Fig.\,\ref{fig:L1-norm} bottom left panel). At the maximum resolution
employed ($n_x=3200$) the errors made with the HLLC solver are
$\sim 26 \%$ smaller than the ones made with the HLL solver. For the
ST3 test, the computational cost to obtain the solution employing the
HLLC solver is $\sim 13\%$ larger than using the HLL solver (see
Tab.\,\ref{tab:ct}).


\subsection{Shock Tube Problem 4 (ST4)}
\begin{figure*}
  \centering
  \scalebox{0.725}{\input{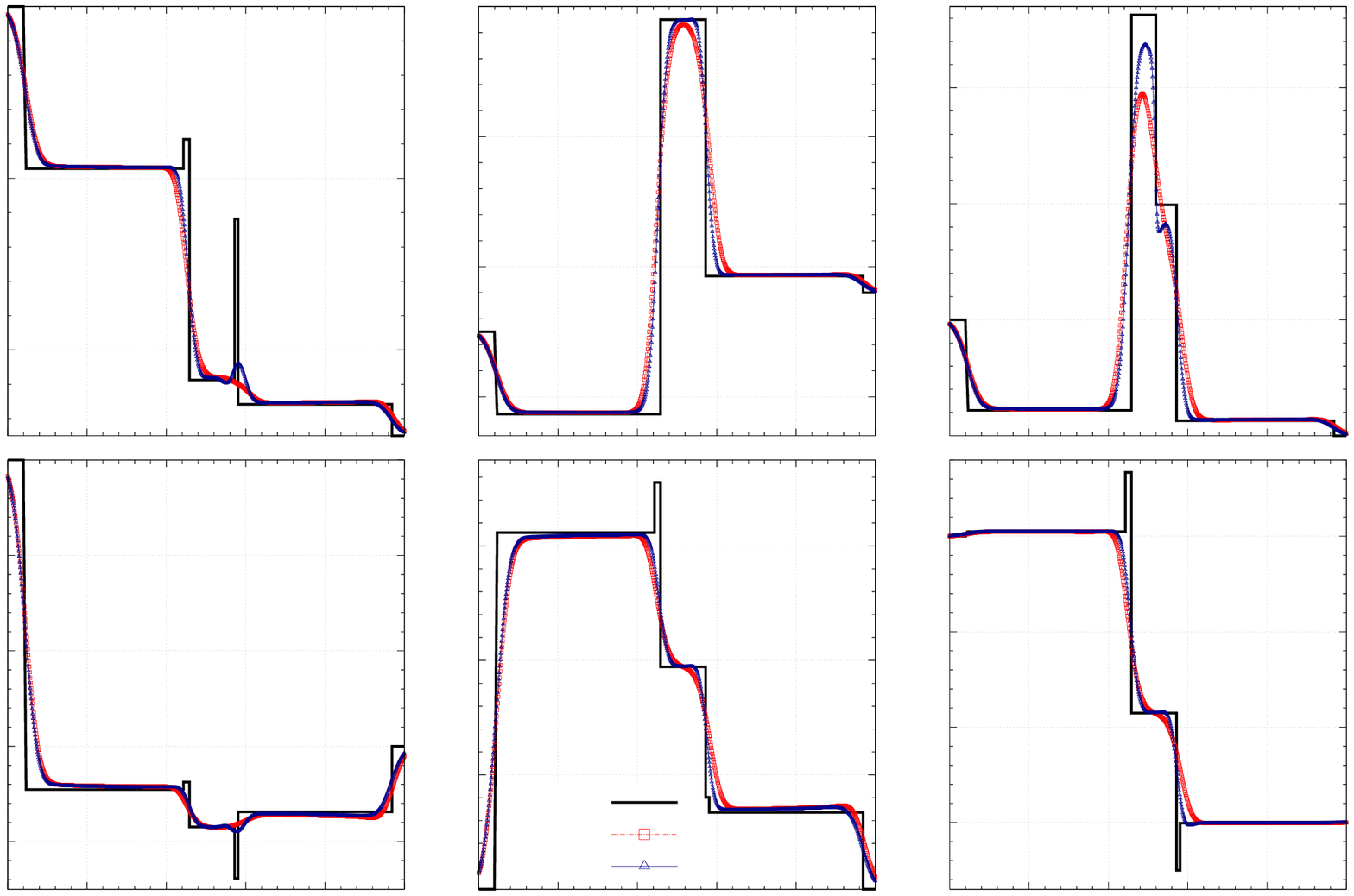}}
  \caption{\textbf{ST4:} Numerical results for the generic Alfv\'en
    test problem ST4, after $t=0.4$ on a grid with 800 numerical
    zones. Upper panels (from left to right): magnetic field ($B_y$
    component), thermal pressure ($p_{\rm g}$), rest-mass density
    ($\rho$). Lower panels (from left to right): electric field ($E_z$
    component), velocity ($v_x$ and $v_y$ components).}
  \label{fig:ST4}
\end{figure*}

\begin{figure*}
  \centering
  \scalebox{0.725}{\input{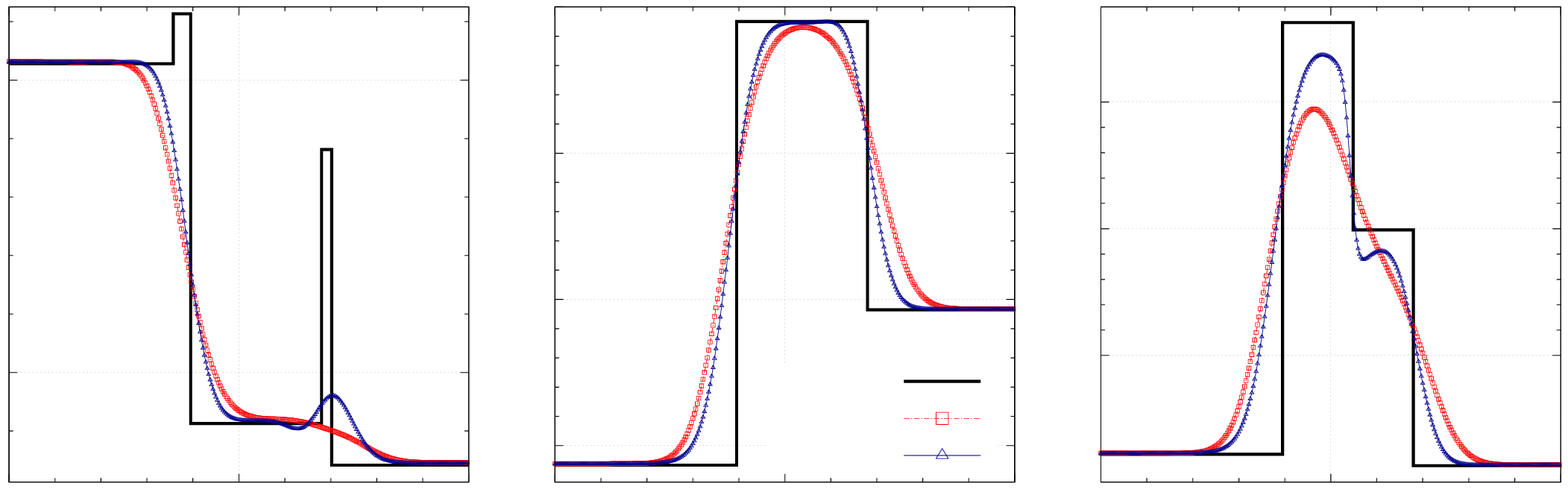}}
  \caption{\textbf{ST4:} Enlargement of the central region of
    Fig.\,\ref{fig:ST4}. From left to right: magnetic field ($B_y$
    component), thermal pressure ($p_{\rm g}$) and rest-mass density
    ($\rho$).}
  \label{fig:ST4_zoom}
\end{figure*}

\cite{Giacomazzo_Rezzolla:2006} refer to this Riemann problem as
``Generic Alfv\'en'' test, which is very challenging for ideal RMHD as
well as RRMHD codes, as it encompasses all seven possible waves in the
Riemann fan it develops. It has been adopted as a benchmark by, e.g.,
\cite{Mignone_etal:2009} and \cite{Anton_etal:2010}. For this setup,
the initial discontinuity results into a contact discontinuity which
separates a fast rarefaction wave (at $x\simeq 0.05$), a rotational
wave (at $x \simeq 0.44$), and a slow shock (at $x \simeq 0.46$), from
a slow shock (at $x \simeq 0.56$), an Alfv\'{e}n wave (at
$x \simeq 0.57$) and a fast shock (at $x \simeq 0.97$). The exact
solution of this test, together with the results at $t=0.4$ for the
HLLC and HLL solvers, are shown in Fig.\,\ref{fig:ST4}. The finest
structures in this test, associated with the rotational
discontinuities travelling very close to the slow shocks, are hardly
resolved by the $1^{\rm st}$-order scheme using any of the approximate
Riemann solvers employed in this paper at the working resolution of
800 uniform numerical zones (Fig.\,\ref{fig:ST4_zoom}).

The L1-norm errors of this test (Fig.\,\ref{fig:L1-norm} bottom mid
panel) show the same qualitative trend than those of the ST2 test
(Fig.\,\ref{fig:L1-norm} upper right panel). The HLLC numerical
solution yields L1-norm errors $\sim 13\%$ smaller than the HLLC
counterpart almost independently of the numerical resolution. For
  the ST4 test, the computational cost to obtain the solution
  employing the HLLC solver is $\sim 18\%$ larger than using the HLL
  solver (Tab.\,\ref{tab:ct}). This represents the largest
  computational overhead of all the 1D tests presented in this
  section. The larger computing time is explained in terms of the
  challenging nature of the ST4 test, that develops several regions
  where very fine structures are attempted to be resolved by the HLLC
  solver (while they are completely smeared out by the HLL solver).

\subsection{Shock Tube Problem 5 (ST5)}

\begin{figure*}
  \centering
  \scalebox{0.725}{\input{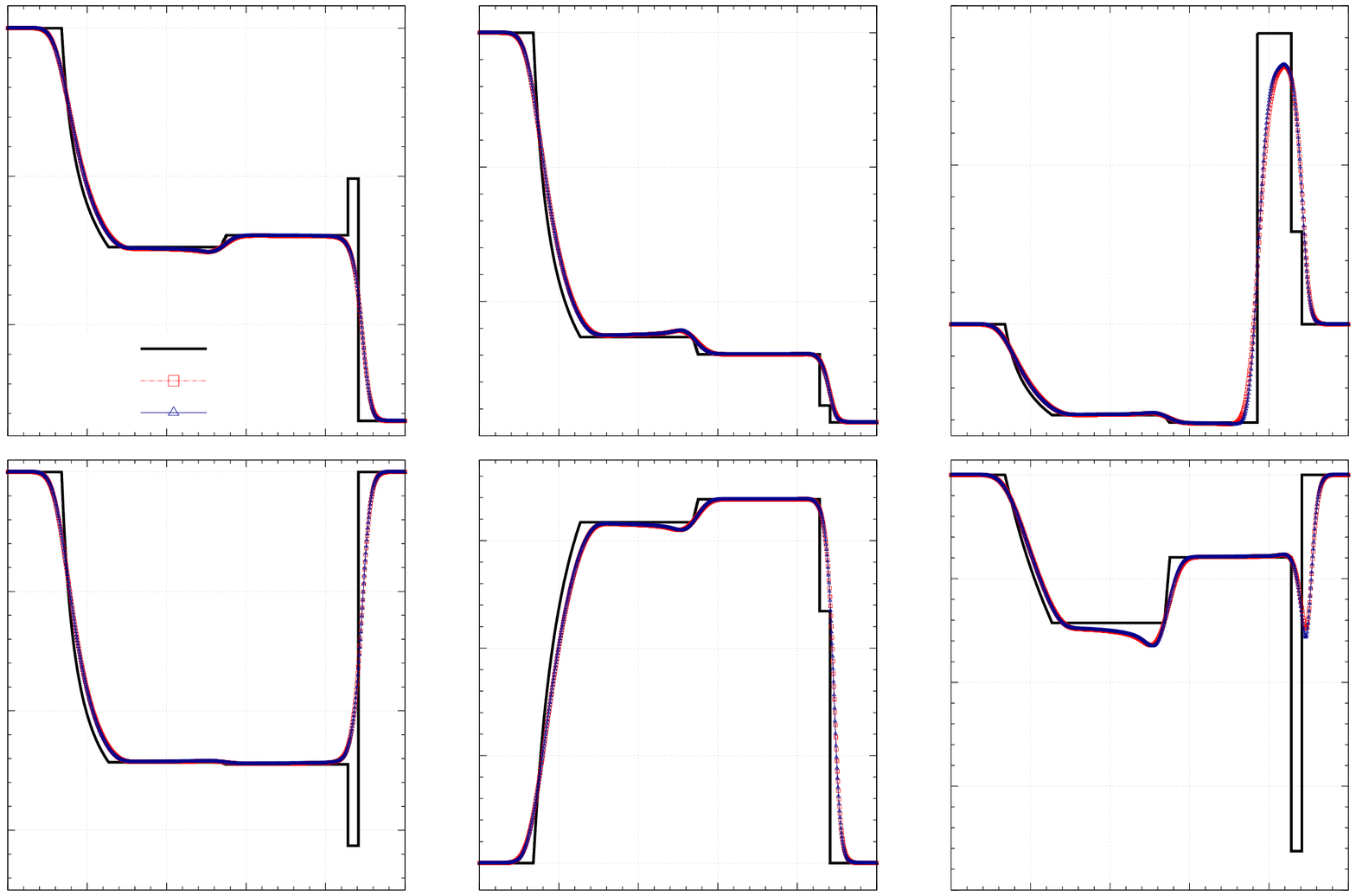}}
  \caption{\textbf{ST5:} Numerical and exact solutions for the ST5
    test at $t=0.4$. Upper panel: magnetic field ($B_y$ component),
    thermal pressure ($p_{\rm g}$), rest-mass density ($\rho$). Lower
    panel: electric field ($E_z$ component), velocity ($v_x$ and $v_y$
    components).}
  \label{fig:ST5}
\end{figure*}
\begin{figure*}
  \centering
  \scalebox{0.725}{\input{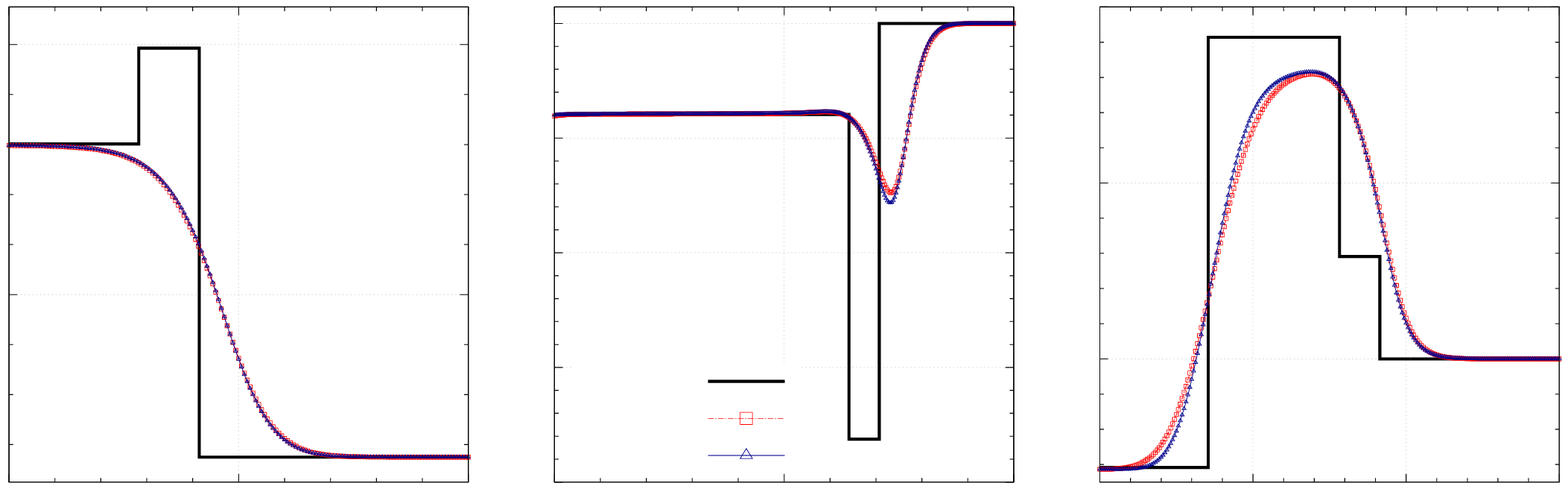}}
  \caption{\textbf{ST5:} Magnification of a region of interest in
    Fig.\,\ref{fig:ST5}. From left to right, magnetic field ($B_y$
    component), velocity ($v_y$ component) and rest-mass density
    ($\rho$).}
  \label{fig:ST5_zoom}
\end{figure*}

The relativistic blast wave test problem with a moderate initial
pressure difference was proposed by \citep{Balsara:2001}. The Riemann
problem develops a left-going fast rarefaction wave (at
$x \simeq 0.16$) and slow rarefaction fan (at $x \simeq 0.53$) are
separated by a contact discontinuity (at $x \simeq 0.76$) from two
right-going shocks waves, a slow (at $x \simeq 0.86$) and fast (at
$x \simeq 0.9$) one. The overal structure of the solution can be found
in Fig.\,\ref{fig:ST5}, where it is evident the inability of either
HLLC or HLL to properly capture the finest structure left to the
right-going slow shock wave. The contact wave is slightly better
resolved with the HLLC solver than with the HLL solver, as can be seen
in Fig.\,\ref{fig:ST5_zoom}. The marginally better performance of the
HLLC solver at all the resolutions considered is quantified in terms
of the L1-norm errors of the rest-mass density
(Fig.\,\ref{fig:L1-norm} bottom right panel). The proximity of the
L1-norm errors obtained with the HLLC and HLL solver is due to the
fact that the contact discontinuity in this test is moving faster than
in any other 1D tests shown in this section. This reduces the
sharpness with which the contact wave is captured by the HLLC scheme,
as discussed for the CW2 test in Sec.\,\ref{sec:CW}.

Finally, we have measured a mean computational overhead of
  $\sim 15\%$ using the HLLC solver in the ST5 test with respect to
  the corresponding models employing the HLL solver
  (Tab.\,\ref{tab:ct}).

\subsection{Resistive rotor (RR)}
\label{sec:rotor}
\begin{figure*}
  \centering
\includegraphics[width=1.0\linewidth]{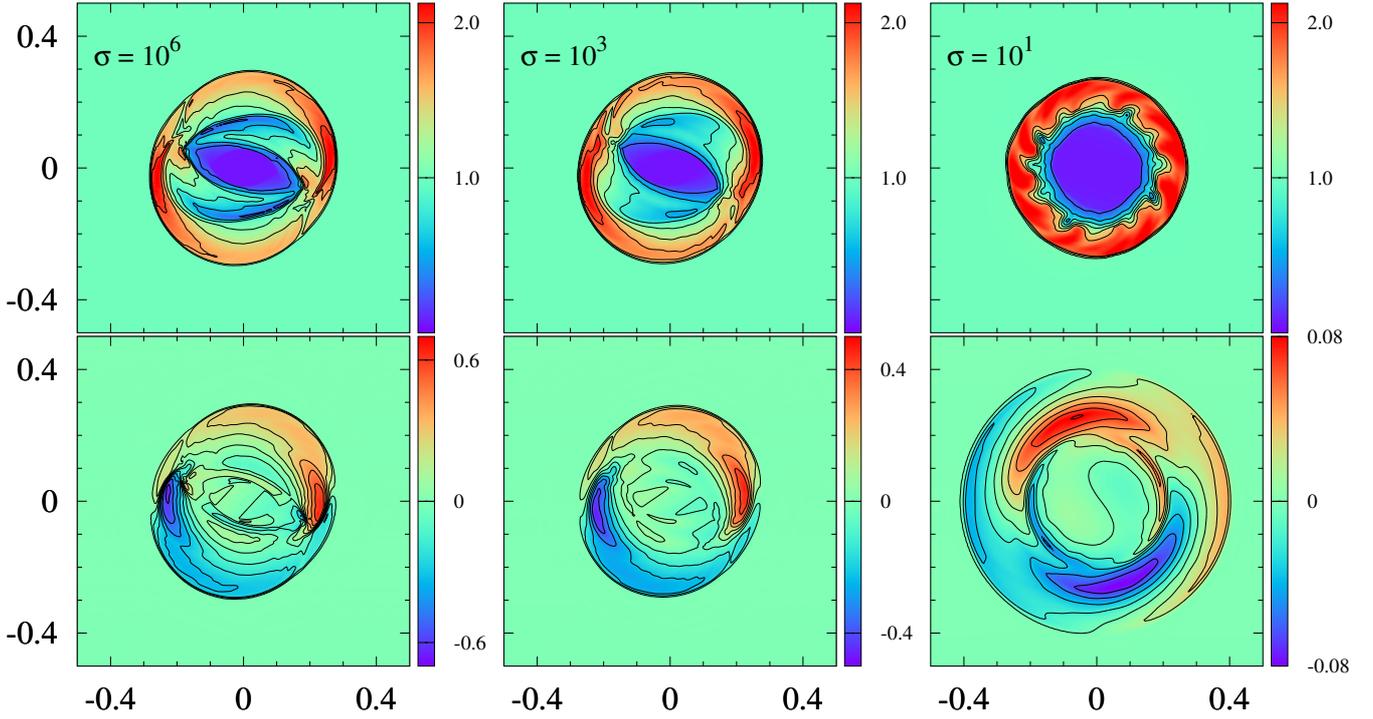} 
\caption{Resistive Rotor computed with a second order MIRK scheme, MP5
  intercell reconstruction and a grid of $300\times 300$ zones evolved
  up to a time $t=0.3$. Upper panels: snapshots of the gas
  pressure. Lower panels: snapshots of $E_z$. We show the evolution
  for different values of the conductivity: the nearly ideal case
  $\sigma = 10^6$ (left panels), an intermediate or semi-resistive
  case $\sigma = 10^3$ (central panels) and a resistive case
  $\sigma = 10$ (right panels).}
  \label{fig:RR}
\end{figure*}

The resistive rotor \citep[see,
e.g.,][]{Dumbser_Zanotti:2009,Bucciantini_DelZanna:2013} is of
  interest not only as a calibration test for resistive as well as
  ideal MHD numerical multidimensional codes, but also because of its
  connection to the problem of angular momentum loss through torsional
  Alfv\'en waves in star formation
  \citep{Mouschovias_Paleologou:1980}. It consists of an initial 2D
state where in a region of radius $r \le 0.1$ around the domain
center, the density is $\rho = 10$, and the fluid rotates with
constant angular velocity $\Omega = 8.5$. Outside this region
($r>0.1$) the medium at rest is uniform ($\rho=1$). Both the pressure
($p_{\rm g}=1$) and the magnetic field $\mathbf{B} = (1,0,0)$ are
uniform in the whole computational domain, a unit square covering the
range $-0.5\le x\le 0.5$, $-0.5\le y \le 0.5$. The initial electric
field is set like in ideal RMHD, and the adiabatic index is
$\gamma = 4/3$. Figure~\ref{fig:RR} shows snapshots of the gas
pressure $p$ and of the electric field component $E_z$ at $t=0.3$, in
different conductivity regimes (cases with $\sigma=10^6$, $10^3$ and
10 are considered). The model was obtained using the second order MIRK
scheme, the MP5 intercell reconstruction, with a CFL factor
$\CFL= 0.1$, the HLLC approximate Riemann solver and a grid of
$300$
zones per dimension.

\begin{figure*}
    \centering
    \scalebox{0.6}{\input{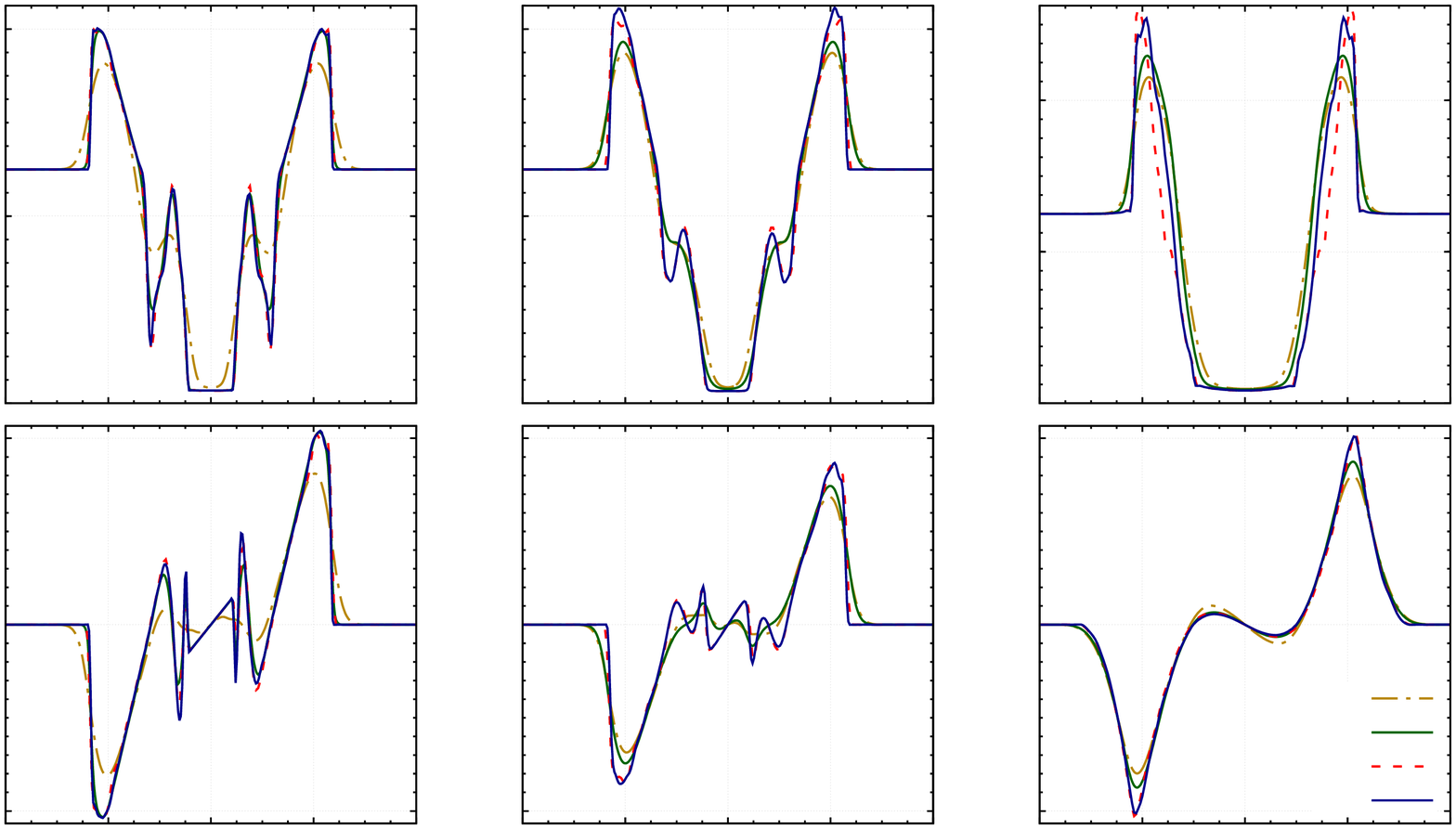}}
    \caption{Profile along the $x=0$ axis of the gas pressure (upper
      panels) and $E_z$ (lower panels) for different values of the
      conductivity (labeled in each panel) at $t=0.3$. Dash-dotted
      orange and solid green lines show the $1^{\rm st}$-order schemes
      employing HLL and HLLC solvers, respectively. Dashed red and
      solid blue lines show the cases in which we combine the
      $5^{\rm th}$-order intercell reconstruction (MP5) with either
      the HLL solver or the HLLC solver, respectively.}
  \label{fig:RR_HLLC_compare}
\end{figure*}

Our results agree relatively well with those of
\cite{Dumbser_Zanotti:2009} and \cite{Bucciantini_DelZanna:2013} for
intermediate ($\sigma=10^3$) or almost ideal ($\sigma=10^6$)
regimes. In the resistive regime ($\sigma=10$), the differences are
more obvious though. The interface between the high-density, rotating
central cylinder and the external medium develops small amplitude
instabilities in our case (visible in the upper right panel of
Fig.\,\ref{fig:RR}). Also the two dimensional distribution of $E_z$
(Fig.\,\ref{fig:RR} lower right panel) {\em seems} more circularly
symmetric in our case. We note, however that the vertical scale of
\cite{Bucciantini_DelZanna:2013} is compressed with respect to the
horizontal one in their Fig.\,3. Hence all the structures look more
oblate than what they actually are in our figures, where the vertical
and horizontal scales are isotropic.  The differences beween our
approach and those of \cite{Dumbser_Zanotti:2009} and
\cite{Bucciantini_DelZanna:2013} likely stem from the distinct
numerical methodologies employed. For instance,
\cite{Bucciantini_DelZanna:2013} employ a {\em fully constrained}
scheme, i.e., they do not have an explicit equation for the charge
conservation as we do, but instead impose $\nabla\cdot
\mathbf{E}=q$. Furthermore, they enforce the magnetic solenoidal
constraint, $\nabla\cdot \mathbf{B}=0$ employing staggered grids via
the upwind constrained transport method \citep{DelZanna_etal:2003},
while we resort to the hyperbolic divergence cleaning method of
\cite{Dedner_etal:2002}. However, we find that the spatial
reconstruction is one of the most relevant differences to explain the
discrepancies in the low-$\sigma$ regime. In such a resistive regime
the wave structure changes drastically, with a faster expanding
electric field, which is no longer inductive (see also the transverse
profile of the solution along the $x=0$ axis in
Fig.\,\ref{fig:RR_HLLC_compare} lower panels).  We have employed the
$5^{\rm th}$-order accurate reconstruction scheme MP5, while
\cite{Bucciantini_DelZanna:2013} resort to a $3^{\rm rd}$-order
accurate Central Essentially Non-Oscillatory spatial reconstruction
with a Monotonised Center (MC) limiter. We have repeated the resistive
rotor test reducing the order of the spatial reconstruction. For that
we have employed a $2^{\rm nd}$-order MC intercell reconstruction,
finding that the differences with respect to
\cite{Bucciantini_DelZanna:2013} for the case $\sigma=10$ are
significantly reduced.  We especially observe a reduction of the
perturbations in the pressure at the transition layer between the
central, rotating cylinder and the initially static outer medium. Our
results hint towards an excessive dissipation of the algorithms, both
of \cite{Bucciantini_DelZanna:2013} and of ours, for low-order spatial
reconstructions. The differences in the resistive regime with respect
to \cite{Dumbser_Zanotti:2009} can also be attributed to the usage of
a lower order scheme ($P_0P_2$ in their case) compared to
ours. Figure~\ref{fig:RR_HLLC_compare} shows a more quantitative
comparison among the HLLC and HLL solvers with different spatial
reconstructions. The smaller numerical resistivity of the HLLC solver
stands out in the nearly-ideal regime ($\sigma=10^6$;
Fig.\,\ref{fig:RR_HLLC_compare} left panels) for tests with a
$1^{\rm st}$-order (Godunov) spatial reconstruction. The HLLC solver
captures very well the sharp, small scale variations in the solution,
which almost overlaps with that of models using the MP5 spatial
reconstruction (compare the green line -HLLC- with the solutions using
MP5 and either HLLC or HLL in the lef panels of
Fig.\,\ref{fig:RR_HLLC_compare}). At intermediate values of the
conductivity ($\sigma=10^3$) the solutions using the HLL and HLLC
solvers are nearly overlapping, and the $1^{\rm st}$-order schemes
smear out the torsional Alfv\'en waves (small scale structure)
flanking the central core of the rotor
(Fig.\,\ref{fig:RR_HLLC_compare} central panels).


In this 2D test, the models run with the HLLC solver need $\sim 5\%$ larger
  computational time than those computed with the HLL solver.

\subsection{Cylindrical Explosion (CE)}
\label{sec:explosion}
The cylindrical explosion test develops a strong shock propagating
into a magnetically dominated medium. Results for different
cylindrical explosion problems in RMHD
\citep[e.g.,][]{Komissarov:1999,DelZanna_etal:2003,Leismann_etal:2005,Mignone_Bodo:2006,Marti:2015}
as well as in the RRMHD regime
\citep[e.g.,][]{Komissarov_etal:2007,Palenzuela_etal:2009} have been
published. We have chosen a setup similar to that in
\cite{Palenzuela_etal:2009}.  It consists of a square covering the
range $-6\le x\le 6$, $-0.5\le y \le 0.5$, having a central circular
region with a radius $r=\sqrt{x^2+y^2}\le 0.8$, where the gas pressure
($p_{\rm g}=1$) and the rest-mass density $\rho=0.01$ are higher than
elsewhere ($p_{\rm g}=\rho=0.001$; $r >1$). We note that these values
of the thermal pressure and of the rest-mass density are larger than
the {\em standard} ones in RMHD benchmarks, where
$p_{\rm g}=3\times 10^{-5}$ and $\rho=10^{-4}$ for $r >1$ are
adopted. Our RRMHD code is unable to handle such extreme conditions,
where the magnetisation of the outer medium is 50 times larger than
assumed here, unless a prohibitively small CFL factor or extremely
fine grids are employed. The central region is continuously connected
with the surroundings using an exponentially decreasing pressure and
density in the region $0.8\le r\le 1$. Everywhere in the computational
domain, the magnetic field,
$\mathbf{B} = (0.1, 0,0)^T$,\footnote{\cite{Palenzuela_etal:2009}
  employ a weaker magnetic field $\mathbf{B} = (0.05, 0,0)^T$ in this
  test.} is uniform, $\mathbf{v}=(0,0,0)^T$ and the adiabatic index is
$\gamma=4/3$. The initial data are evolved until $t=4$. This test is
used to validate the new resistive code in 2D and in the ideal limit
(a uniform conductivity $\sigma = 10^6$ is set everywhere), in a
situation where strong shocks develop (in contrast to the resistive
rotor shown in Sect.~\ref{sec:rotor}).

\begin{figure*}
  \centering
  \includegraphics[width=1.0\linewidth]{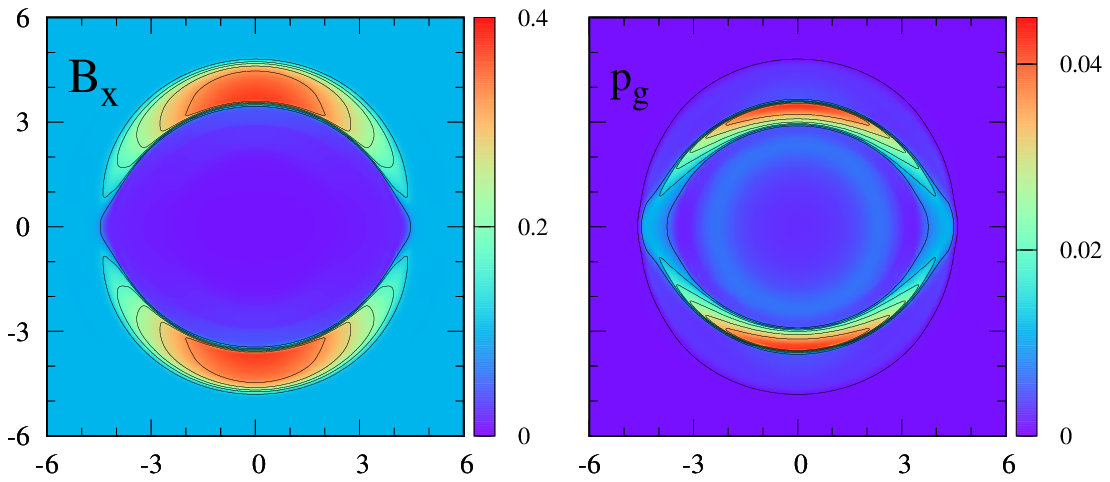} \\
  \hspace{-0.3cm}\scalebox{0.55}{\input{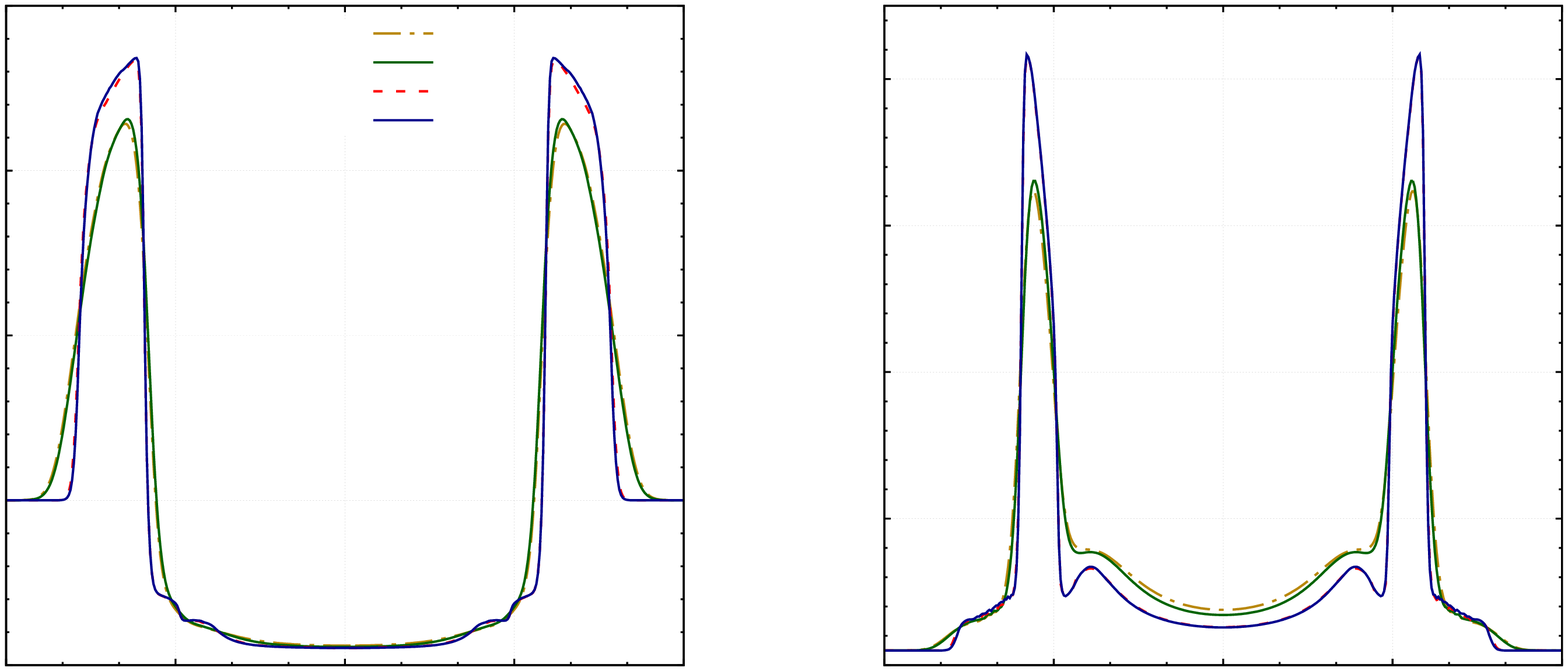}}
  \caption{Upper panel: snapshots at $t=4$, of the magnetic field ($B_x$ component; left) and thermal pressure $p_g$
    (right), for the CE test computed with a uniform grid of $400 \times 400$ cells, CFL factor $\CFL=0.1$, employing
    the IMEX-RK scheme SSP2(332)-LUM and a MCL intercell reconstruction. Lower panel: profiles along $x=0$, for $B_x$
    (left) and $p_g$ (right). Dash-dotted orange and solid green lines show the $1^{\rm st}$-order HLL and HLLC schemes,
    respectively. Dashed red and solid blue lines show the second order case using the MC spatial reconstruction in
    combination with the HLL and the HLLC solvers, respectively.}
  \label{fig:CE}
\end{figure*}

A direct comparison with an analytic solution is not possible in this
case. We note, however, that our results (Fig.~\ref{fig:CE}) compare
fairly well with the those obtained with our ideal RMHD code
\citep{Leismann_etal:2005,Anton_etal:2010}, as well as with the same
setup in \cite{Palenzuela_etal:2009}. As in the latter reference, our
solution is also regular everywhere and similar results can be
obtained with smaller values of the conductivity, namely with
$\sigma\gtrsim 10^4$.

The largest discontinuities in this test are the shocks delimiting the fast expanding shell, whose profiles at
  $x=0$ are displayed in Fig.\,\ref{fig:CE} (bottom panels). Thus, resolving the contact waves does not play a
  major role in the overall dynamics. As a result, the advantage of using the HLLC solver with respect to employing the
  HLL solver is significantly decreased. Indeed, for this test, the HLLC and HLL solutions basically overlap
  independently of the spatial reconstruction employed. This fact is evident in Fig.~\ref{fig:CE} (bottom panels). The
  solutions computed without any spatial reconstruction are nearly coincident, with tiny discrepancies in the pressure
  on the central evacuated area of the domain (see yellow and green lines in Fig.~\ref{fig:CE} bottom panels). Employing
  a $2^{\rm nd}$-order MC reconstruction the HLL and HLLC solutions are almost indistinguishable.

The CE tests performed with the HLLC solver require $\sim 18\%$
  larger computational time than those computed with the HLL solver,
  i.e., in this 2D test the computational overhead of using the HLLC
  solver is a bit larger than for the RR test. However, looking at the
  variations of the overheads displayed in Tab.\,\ref{tab:ct}, we may
  conclude that the computational time needed by the HLLC solver is
  $\sim 5 - 20\%$ larger than with the HLL solver, independently of
  the dimensionality of the problem.
%
\section{Astrophysical applications: relativistic \maa{ideal} tearing modes}
\label{sec:TM}

The purpose of this section is twofold. On the one hand, we
  present an astrophysically relevant application of the newly
  developed numerical method and, on the other hand, we aim to
  calibrate the ability of the HLLC and HLL solvers to properly obtain
  the growth rate of relativistic \maa{ideal} TMs.

The TM instability is a resistive MHD instability that can develop in
current sheets and dissipates magnetic energy into kinetic energy and
subsequently into thermal energy.  TMs disconnect and rejoin magnetic
field lines, thereby changing the topology of the magnetic field. The
linear theory of TMs was extensively studied, in the context of plasma
fusion physics, in a seminal paper of \cite{Furth_etal:1963}.  TMs are
of great relevance in astrophysics, (e.g.~in the magnetopause or
magnetotail of the solar wind, in flares or coronal loops of the Sun,
and in the flares of the Crab pulsar;
cf.\,\citealt{Priest_Forbes:2000}). 
They have been also suggested to be a terminating agent of the MRI
\citep[][but see \citealt{Rembiasz_etal:2016b} who observed an MRI
termination by the Kelvin-Helmholtz instability in their 3D MRI
simulations]{Balbus_Hawley:1991,Latter_etal:2009,Pessah:2010}.

Following \cite{DelZanna_etal:2016}, we simulate a relativistic
\maa{ideal} TM in a 2D domain of $[-20a, 20a] \times [0, L_y]$, where
$a = 0.01 L$, $L_y = 2 \pi / k$, and we further set $L = 1$ and
$k = 12$.  \maa{The term \emph{ideal} TM was introduced by}
\cite{Pucci_Velli:2014} \maa{who pointed out that current sheets with
  a thickness $a = S^{-1/3}L$ (where $S$ is the Lunquist number
  defined below) are unstable agains a TM growing on an Alfv\'en
  (\emph{ideal}) timescale in classical resistive MHD.}
\cite{Landi_etal:2015} \maa{confirmed that numerically with 2D
  compressible classical MHD simulations.}  We use copying and
periodic boundary conditions in the directions $x$ and $y$,
respectively, for all variables but $B_x$. The boundary values of the
latter variable in the $x$ direction are computed from the solenoidal
constraint $\nabla \cdot \mathbf{B}=0$.
To  trigger the TM
  instability, we perturb the initial background magnetic field 
\begin{align}
  \label{eq:tm_bg}
  B_{0y} &= B_0 \tanh ( x / a), \\
  B_{0z} &= B_0 \ {\rm sech} ( x / a),
\end{align}
with
\begin{align}
  \label{eq:tm_perts}
  B_{1x} &= \epsilon B_0  \cos ( k y )  \ {\rm sech} ( x / a), \\
  B_{1y} &=  \epsilon k^{-1}  B_0  \sin( k y) \tanh ( x/a)   \ {\rm sech} ( x / a).
\end{align}
We set $B_0 = \rho_0 = 1$, $p_0 = 0.5$, $\sigma = 2 \times 10^6$, and
$\epsilon = 10^{-4}$, so that the magnetisation is
$\sigma_m = B_0^2 / \rho_0 = 1$, the ratio of magnetic-to-thermal
pressure becomes $\beta_0 = B_0^2/ (2p_0) = 1 $, and the resulting
Alfv\'en speed and Lunquist number are $c_A = 0.5$ and
$S \equiv L c_A \sigma = 10^6$, respectively.  For this Lundquist
number, \cite{DelZanna_etal:2016} used a resolution of
$2048 \times 512$ zones. This choice was driven by the fact that for
the default resolution chosen by those authors ($1024 \times 512$),
numerical resistivity was higher than the physical one and strongly
affected their simulation results.  However, in our studies, we used
much more moderate resolutions of $512 \times 32$ and $1024 \times 32$
zones for the following reasons. First, because we employ an ultra
high-order spatial reconstruction scheme of the $9^{\rm th}$-order
\citep[MP9;][]{Suresh_Huynh_1997} whose numerical dissipation is much
lower than that of other lower order schemes \citep[see the extensive
studies of][]{Rembiasz_etal:2017}.  Second, as we are only interested
in the linear phase of the TM instability, we can use a much lower
resolution in the $y$-direction where all perturbed quantities exhibit
a (co-)sinusoidal variation.  Hence, the characteristic length of the
system \citep{Rembiasz_etal:2017} in this direction is equal to the
box length, $L_y$, and it can be very well resolved with $32$ zones
using the MP9 scheme.
\begin{figure*}
  \centering
  \scalebox{0.8}{\input{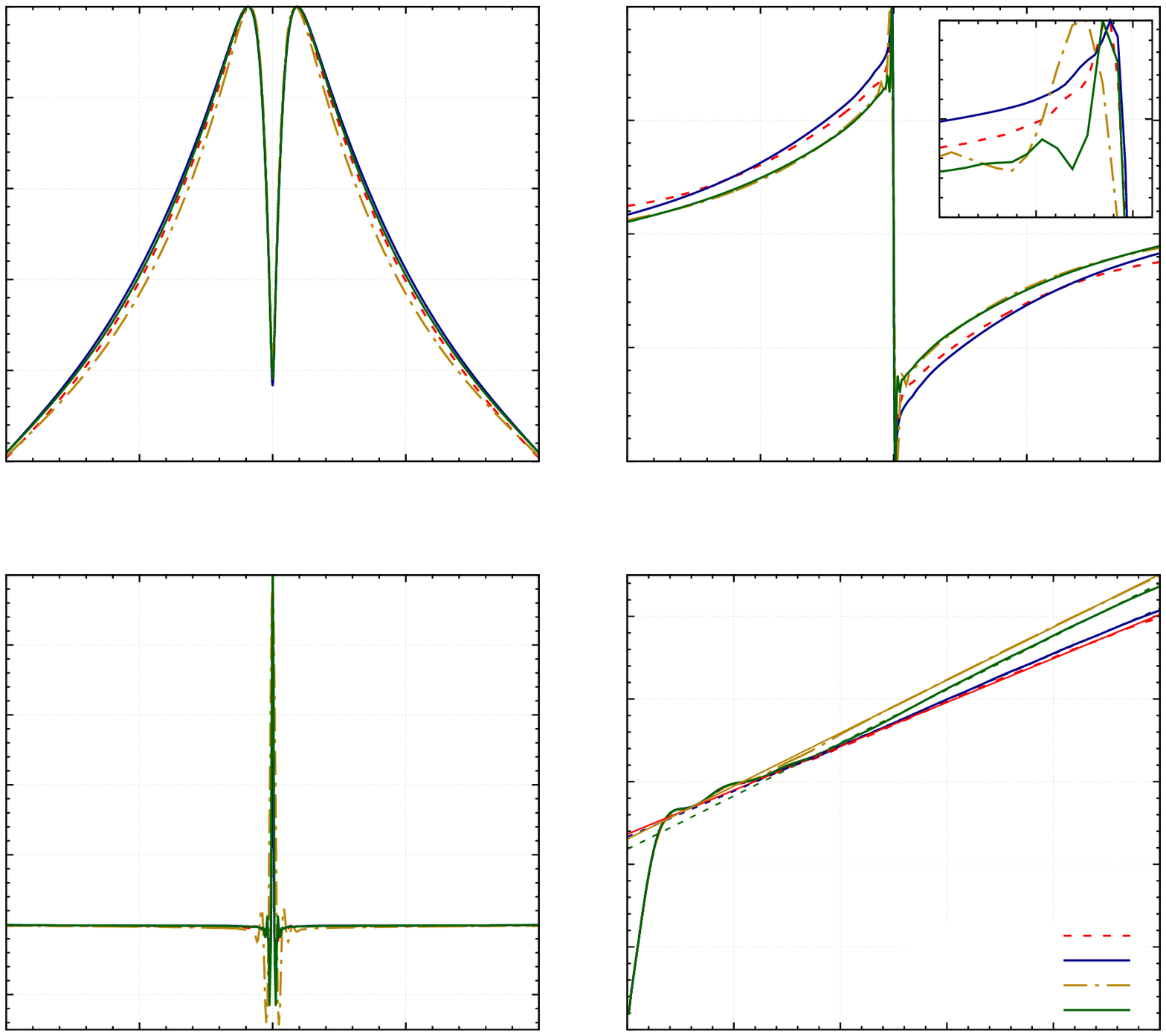}} \\
  \caption{ %
    2D simulations of TMs performed with the HLL and HLLC Riemann
    solvers, MP9 reconstruction scheme and resolution of
    $512\times 32$ and $1024\times 32$ zones.  All quantities (but in
    bottom right panel) are presented at $t = 10$ and normalised for a
    better comparison.  \emph{Upper left}: $x-$component of magnetic
    field ($B_x$) at $y = 0$.  \emph{Upper right}: $x-$component of
    velocity ($v_x$) at $y = 3L_y/4$.  \emph{Bottom left}:
    $y-$component of velocity ($v_y$) at $y = 0$.  \emph{Bottom
      right}: Time evolution of $\ln (\int B_x^2 \,\mathrm{d}S )$. The
    TM
    growth rate is determined from a linear fit (solid and dashed
    lines of the corresponding colours for the HLL and HLLC solvers, respectively)
    to this quantity for
    $t\in [4,10]$ (see Tab.~\ref{tab:tm}).%
  }
  \label{fig:tm}
\end{figure*}
\begin{table*}
\centering 
\caption{TM growth rate determined from a linear
  fit of $\ln{\cal B}$ (Eq.\,\ref{eq:fitformula}) in the time interval $t\in [4,10]$  (see bottom right panel of Fig.\,\ref{fig:tm})
  in  simulations performed with the HLLC (second column) and HLL (third
  column) approximate Riemann solvers and the MP9 spatial reconstruction scheme.}
\begin{tabular}{ ccc }
\multicolumn{3}{c}{\textbf{Growth rate}}                                             \\  \hline \hline
resolution                         &          HLLC            &        HLL           \\ \hline
$512  \times 32$  &   $0.322 $    &  $0.321$   \\ \hline
$1024 \times 32$ &   $0.276 $    &  $0.265$   \\ \hline
\end{tabular}
\label{tab:tm}
\end{table*}
A possible tracer for the growth of the TM instability is the induced
growth of the magnetic field component $B_x$, which after the initial
transient phase is assumed to grow as
$B_x(t)=B_{1x} e^{\gammatm t }$, where $B_{1x} $ is a time independent
eigenfunction of the TM. In order to have a positively defined global
quantity we compute the integrated value on the whole computational
domain of $B_x^2$, i.e., ${\cal B}:=\int B_x^2(t) \,
\mathrm{d}S$. Then, we take the logarithm of ${\cal B}$,
\begin{equation}
  \ln{\cal B} = 2\gammatm t  +  \ln \left(\int B_{1x}^2 \, \mathrm{d}S \right),
\label{eq:fitformula}
\end{equation} 
and obtain $\gammatm$ from the slope of the linear fit $\ln{\cal B}$
vs $t$ \footnote{We note that \cite{DelZanna_etal:2016} employs a
  different variable to compute $\gammatm$, namely
  $\ln( \mathrm{ \max{( B_x )}})$.} in the time interval
$t\in [4,10]$.

Our simulation results (Fig.\,\ref{fig:tm}
and Tab.\,\ref{tab:tm}) are very similar to those obtained by
\cite{DelZanna_etal:2016} (Figs.~1, 3 and 4, therein).  The TM
instability sets in after $t \approx 2$ and its growth rate (bottom
right panel of Fig.~\ref{fig:tm}) is close to
$\gammatm = 0.3$, which \cite{DelZanna_etal:2016} obtained both
analytically as well as with their numerical code for solving the
  linearised MHD equations.  In simulations performed with the lower
resolution, i.e., of $512\times 32$ zones, employing both Riemann
solvers, the TM growth-rate is very similar but higher than
theoretically expected, i.e., $\gammatm \approx 0.32$. We attribute
this discrepancy to numerical resistivity.  In the simulations
performed with $1024 \times 32$ zones, the TM growth-rate is lower
than theoretically expected, i.e.\ $\gammatm \approx 0.27$. This can
be explained by the resistive dissipation of the background magnetic
field \citep[as pointed out by][who obtained the same
value]{DelZanna_etal:2016} as well as by numerical viscosity present
in the simulations, since viscosity is known to reduce the TM
  growth rate \citep{Furth_etal:1963,Rembiasz_etal:2017}.
The $x-$components of the magnetic field
($B_x$) 
(see the upper left panel of
Fig.\,\ref{fig:tm}) 
are similar in all four simulations, however, there are visible
differences in the $x-$ and $y-$components of the
velocity. 
In the simulation performed with the HLL solver and the resolution of
$512 \times 32$ zones, the (characteristic for the TM instability)
velocity peaks of the $v_x$ component (Fig.\,\ref{fig:tm}, upper
  right panel) are located farther away from
$x = 0$ than in the other simulations.  We attribute this difference
to a higher numerical viscosity (and resistivity) of the HLL solver,
as the distance between these peaks is proportional to certain (not
necessarily equal) powers of viscosity and resistivity \citep[see][
for a detailed discussion for a different TM
setup]{Rembiasz_etal:2017}.  However, it is not necessary
  invoking the differences in numerical resistivity and viscosity
  (whose accurate knowledge requires a very careful calibration of the
  numerical method) between the two solvers employed to assess the
  superiority of the HLLC solver. Instead, we may compare solutions
  computed at the two different resolutions used in these series of
  tests. Looking at the profiles of both $v_x$ and $v_y$, their
  resemblance in the case of models run with the HLLC solver is
  greater than when using the HLL solver. Since the higher resolution
  models are closer to the actual solution of the problem, this means
  that tests conducted with the HLLC solver are closer to a converged
  state than the same models run with the HLL solver.

\section{Conclusions}
\label{sec:conclusions}

We have developed a new HLLC approximate Riemann solver for the
augmented RRMHD system of equations. The extra equations of the system
are used to control the violation of the solenoidal magnetic field
constraint and enforce charge conservation to truncation error. The
new solver captures exactly isolated stationary contact discontinuities,
improving on the single state HLL solver, which in RRMHD reduces to a
Local-Lax-Friedrich (global scheme). The new HLLC solver does not need
to distinguish between the cases in which the magnetic field
perpendicular to a discontinuity is zero or not and thus it does not
suffer from any pathological singularity when the component of
magnetic field normal to a zone interface approaches zero. Several
test problems in 1D and 2D show that the HLLC scheme always displays
smaller numerical diffusion than the HLL approximate Riemann solver.
The computational overhead with respect to the simpler HLL solver is
very modest and its implementation in existing RRMH codes is
straightforward.

The new solver has shown its capability to resolve strong shocks in 2D
tests and a good behaviour in different conductivity regimes. For most
of the numerical experiments considered, the results are insensitive
to the exact value of the conductivity when $\sigma\gtrsim 10^6$. We
take this as an indication of the fact that the intrinsic numerical
resistivity of our algorithm is $\lesssim 10^{-6}$. Also, this result
justifies our choice of a default conductivity ($\sigma=10^6$) to
adress the ideal RMHD regime in most of the numerical benchmarks we
have conducted.

We find that the models run with the HLLC solver are $\sim 5\% - 20\%$ more computationally expensive than the
  corresponding counterparts employing the HLL solver (Tab.\,\ref{tab:ct}). This results hold for both 1D and 2D
  simulations. The variations in the computational time are closely related to the different number of iterations
  necessary to solve numerically the quadratic equation (Eq.\,\ref{eq:lambda*}). In view of the fact that the L1-norm
  errors are systematically smaller employing the HLLC solver than the HLL solver, and also considering the small
  computational overhead that the HLLC solver introduces, we conclude that the new HLLC solver is a viable alternative
  to the very broadly used HLL solver. 

In our applications with more astrophysical interest, the HLLC Riemann solver also proved to be superior to the HLL
  solver in 2D simulations of the TM instability. The combination of high-oder spatial reconstruction and the HLLC
  solver helps to accurately estimate the growth rate of the TM instability. We plan to exploit this fact to explore in
  greater detail the physics of relativistic TMs in the future.

\section*{Acknowledgements}

We acknowledge support from the European Research Council (grant
CAMAP-259276) \maa{as well as} from grants AYA2015-66899-C2-1-P and
PROMETEOII/2014-069.  S.\,M-A.\, acknowledges financial support from
COLCIENCIAS conv.~679.  The computations have been performed at the
Servei d'Inform\`atica of the University of Valencia.  \maa{We also
  thank the anonymous referee whose valuable comments and suggestions
  allowed us to improve the quality of this manuscript.}

\appendix
\section{Alternative HLLC solvers}
\label{sec:HLLCa}

\maa{Here, we explore alternatives to the assumption made in
  Sec.~\ref{sec:HLLC1} (Eq.\,\ref{eq:continuity_ContactWave2}) on the
  continuity of the electromagnetic variables. For that, we may
  generalise the approach of \cite{Mignone_Tzeferacos:2010} for RRMHD
  in combination with the GLM method. When solving a one dimensional
  Riemann problem at a zone interface (e.g., in the $x-$direction as
  we are assuming), the following $2\times 2$ linear hyperbolic
  sub-systems arise:
\begin{eqnarray}
\begin{cases}
\begin{aligned}
\partial_t B_x = & -\partial_x \phi, \\
\partial_t \phi = & -\partial_x B_x ,\\
\end{aligned}
\label{eq:Bx-phi}
\end{cases} \\
\begin{cases}
\begin{aligned}
\partial_t E_x = & -\partial_x \psi ,\\
\partial_t \psi = & -\partial_x E_x ,\\
\end{aligned}
\label{eq:Ex-psi}
\end{cases}\\
\begin{cases}
\begin{aligned}
\partial_t B_y = & +\partial_x E_z ,\\
\partial_t E_z = & +\partial_x B_y ,\\
\end{aligned}
\label{eq:By-Ez}
\end{cases}\\
\begin{cases}
\begin{aligned}
\partial_t B_z = & -\partial_x E_y ,\\
\partial_t E_y = & -\partial_x B_z .\\
\end{aligned}
\label{eq:Bz-Ey}
\end{cases}
\end{eqnarray}
We note that in the systems (\ref{eq:By-Ez}) and (\ref{eq:Bz-Ey}) the
electric current and the source terms for the GLM scalar potentials
are not included since they are treated implicitly by our time
integration schemes (either MIRK or RKIMEX).  For generic pairs of
left and right states $(B_{x,l},\phi_l)$, $(B_{x,r},\phi_r)$;
$(E_{x,l},\psi_l)$, $(E_{x,r},\psi_r)$; $(B_{y,l},E_{z,l})$,
$(B_{y,r},E_{z,r})$, and $(B_{z,l},E_{y,l})$, $(B_{z,r},E_{y,r})$, the
Godunov flux of the systems (\ref{eq:Bx-phi})-(\ref{eq:Bz-Ey}) can be exactly
computed  as
\begin{eqnarray}
\begin{cases}
\begin{aligned}
\hat{B}_x = & \frac{B_{x,l}+B_{x,r}-(\phi_r-\phi_l)}{2}, \\
\hat{\phi} = & \frac{\phi_{x,l}+\phi_{x,r}-(B_{x,r}-B_{x,l})}{2} ,\\
\end{aligned}
\label{eq:Bx-phi-sol}
\end{cases} \\
\begin{cases}
\begin{aligned}
\hat{E}_x = & \frac{E_{x,l}+E_{x,r}-(\psi_r-\psi_l)}{2}, \\
\hat{\psi} = & \frac{\psi_{x,l}+\psi_{x,r}-(E_{x,r}-E_{x,l})}{2} ,\\
\end{aligned}
\label{eq:Ex-psi-sol}
\end{cases}\\
\begin{cases}
\begin{aligned}
\hat{B}_y = & \frac{B_{y,l}+B_{y,r}+(E_{z,r}-E_{z,l})}{2} , \\
\hat{E}_z = & \frac{E_{z,l}+E_{z,r}+(B_{y,r}-B_{y,l})}{2} ,\\
\end{aligned}
\label{eq:By-Ez-sol}
\end{cases}\\
\begin{cases}
\begin{aligned}
\hat{B}_z = & \frac{B_{z,l}+B_{z,r}-(E_{y,r}-E_{y,l})}{2} , \\
\hat{E}_y = & \frac{E_{y,l}+E_{y,r}-(B_{z,r}-B_{z,l})}{2} .\\
\end{aligned}
\label{eq:Bz-Ey-sol}
\end{cases}
\end{eqnarray}
Therefore, we may obtain the solution of the $2\times 2$ linear
Riemann problems separately before using a standard Riemann solver for
the remaining set of one-dimensional equations. The electric and
magnetic field components as well as the scalar potentials $\phi$ and
$\psi$ precomputed with
Eqs.\,(\ref{eq:Bx-phi-sol})-(\ref{eq:Bz-Ey-sol}) enter as constant
parameters in the computation of the numerical Riemann fluxes. Our
choice is to employ an HLLC approximate Riemann solver built as in
Sec.\,\ref{sec:HLLC1}. In this case, the only numerical fluxes that
need to be computed are the ones corresponding to
$(q,D^*,S_x^*,S_y^*,S_z^*,{\cal E}^*)$.}

\maa{We note that the numerical fluxes of the conserved variables
  $(\phi,\psi,B_x,B_y,B_z,E_x,E_y,E_z)$ corresponding to the exact
  solutions written in
  Eqs.\,(\ref{eq:Bx-phi-sol})-(\ref{eq:Bz-Ey-sol}) are
  $(\hat{B}_x,\hat{E}_x,\hat{\phi},-\hat{E}_z,\hat{E}_y,\hat{\psi},\hat{B}_z,-\hat{B}_y)$,
  respectively. It turns out that the expressions for these fluxes
  coincide with those that we obtain in the HLLC solver described in
  the Sec.\,\ref{sec:HLLC1} (Eq.\,\ref{eq:Fluxes*q-B-E}) if we fix
  $\lambda_l=-1$ and $\lambda_r=+1$. Remarkably, the numerical fluxes
  of Eq.\,(\ref{eq:Fluxes*q-B-E}) result from the assumption that the
  conserved variables $(\phi,\psi,B_x,B_y,B_z,E_x,E_y,E_z)$ are
  continuous across the contact wave. Therefore, we do not expect that
  precomputing the exact solutions given in
  (\ref{eq:Bx-phi-sol})-(\ref{eq:Bz-Ey-sol}) and then using them as
  parameters in the rest of the HLLC solver may bring any improvement
  in the numerical solution of the RRMHD equations in comparison with
  employing the full HLLC solver as presented in in
  Sec.~\ref{sec:HLLC1} with $\lambda_l=-1$ and
  $\lambda_r=+1$. Nonetheless, the usage of
  Eqs.\,(\ref{eq:Bx-phi-sol})-(\ref{eq:Bz-Ey-sol}) as parameters for
  the rest of the solver may yield a reduced numerical viscosity of
  the algorithm. The reason is that we are not necessarily bound to
  use values $\lambda_l=-1$ and $\lambda_r=+1$ as the limitting speeds
  for the variables $(q,D^*,S_x^*,S_y^*,S_z^*,{\cal E}^*)$. For this
  reduced set of conserved variables, the eigenvalues of the
  corresponding Jacobian matrix are $\lambda_q=\lambda_{H_0}=v_x$ and
  $\lambda_{H_\pm}$ (Eqs.\,\ref{eq:lambdaq}-\ref{eq:lambdaHpm}), and
  the maximum and minimum signal speeds can be taken as
\begin{equation}
\begin{aligned}
\lambda_l&=\min{\{\lambda_{H_-,l}, \lambda_{H_-,r}\}}, \\
\lambda_r&=\max{\{\lambda_{H_+,l}, \lambda_{H_+,r}\}}.
\end{aligned}
\label{eq:lambdal-lambdar}
\end{equation}
We have tested this {\em hybrid} HLLC solver with all the 1D and 2D
tests presented in Sections \ref{sec:tests} and \ref{sec:TM} and found
that it basically provides the same quantitative results as the solver
delineated in Sec.\,\ref{sec:HLLC1}. Nevertheless, this alternative
solver introduces less numerical viscosity, which manifests itself in
small oscillations arising close to discontinuities in 2D tests,
making it a bit less robust to, e.g. accurately predict the growth
rate of relativistic TMs.}


\bibliographystyle{mnras}
\bibliography{hllc_rrmhd_v3}

%
\bsp	
\label{lastpage}
\end{document}